\documentclass[aps,prx,reprint,groupedaddress,amsmath,amssymb]{revtex4-2}

\usepackage[]{hyperref}
\usepackage{graphicx}
\usepackage{enumitem}
\usepackage{siunitx}

\newcommand{\dd}{{\mathrm{d}}}

\newcommand{\dx}{{\dd x}}
\newcommand{\dX}{{\dd X}}
\newcommand{\dXp}{{\dd X'}}

\newcommand{\boldx}{{\boldsymbol x}}
\newcommand{\boldu}{{\boldsymbol u}}
\newcommand{\boldX}{{\boldsymbol X}}

\newcommand{\boldtheta}{{\boldsymbol \theta}}
\newcommand{\boldnabla}{{\boldsymbol \nabla}}

\newcommand{\bdx}{{\dd\boldx}}

\newcommand{\erf}{{\mathrm{erf}}}

\newcommand{\phiE}{{\phi}}
\newcommand{\phiS}{{\phi_{\mathrm{s}}}}

\newcommand{\FE}{{F_{\mathrm{el}}}}
\newcommand{\FI}{{F_{\mathrm{int}}}}

\newcommand{\FF}{{F_{\mathrm{frac}}}}
\newcommand{\FC}{{F_{\mathrm{con}}}}
\newcommand{\barfE}{{\bar{f}_{\mathrm{el}}}}
\newcommand{\barfI}{{\bar{f}_{\mathrm{int}}}}

\newcommand{\barfF}{{\bar{f}_{\mathrm{frac}}}}
\newcommand{\barfC}{{\bar{f}_{\mathrm{con}}}}

\newcommand{\kBT}{{k_{\mathrm{B}}T}}
\newcommand{\volS}{\nu}

\newcommand{\phiEsym}{{\phi_{\mathrm{el}}}}

\newcommand{\barphiEsym}{{\bar{\phi}_{\mathrm{el}}}}

\newcommand{\barphiS}{{\bar{\phi}_{\mathrm{s}}}}

\newcommand{\wE}{{w_{\mathrm{el}}}}
\newcommand{\wS}{{w_{\mathrm{s}}}}
\newcommand{\QE}{{Q_{\mathrm{el}}}}
\newcommand{\QS}{{Q_{\mathrm{s}}}}

\newcommand{\homoName}{homogeneous phase}

\newcommand{\patName}{patterned phase}

\newcommand{\patNameAbbr}{pattern}
\newcommand{\PatNameAbbr}{Pattern}

\newcommand{\RN}[1]{  \textup{\uppercase\expandafter{\romannumeral#1}}}

\newcommand{\Eqref}[1]{Eq.~\ref{#1}}
\newcommand{\Eqsref}[1]{Eqs.~\ref{#1}}
\newcommand{\figref}[1]{Fig.~\ref{#1}}

\hypersetup{
  colorlinks   = true, 
  urlcolor     = blue, 
  linkcolor    = blue, 
  citecolor   = red 
}

\newcommand{\nocontentsline}[3]{}
\newcommand{\tocless}[2]{\bgroup\let\addcontentsline=\nocontentsline#1{#2}\egroup}

\newcommand{\supportingName}{Appendix}

\begin{document}

\title{Theory of Elastic Microphase Separation}

\author{Yicheng Qiang}
\affiliation{Max Planck Institute for Dynamics and Self-Organization, Am Fa{\ss}berg 17, 37077 G{\"o}ttingen, Germany}
\author{Chengjie Luo}
\affiliation{Max Planck Institute for Dynamics and Self-Organization, Am Fa{\ss}berg 17, 37077 G{\"o}ttingen, Germany}
\author{David Zwicker}
\email[]{david.zwicker@ds.mpg.de}
\affiliation{Max Planck Institute for Dynamics and Self-Organization, Am Fa{\ss}berg 17, 37077 G{\"o}ttingen, Germany}

\begin{abstract}
    Elastic microphase separation refers to equilibrium patterns that form by phase separation in elastic gels. Recent experiments revealed a continuous phase transition from the homogeneous phase to a regularly patterned phase, whose period decreased for stiffer systems. We here propose a model that captures these observations. The model combines a continuous field of the elastic component to describe phase separation with nonlocal elasticity theory to capture the gel's microstructure. Analytical approximations unveil that the pattern period is determined by the geometric mean between the elasto-capillary length and a microscopic length scale of the gel. Our theory highlights the importance of nonlocal elasticity in soft matter systems, reveals the mechanism of elastic microphase separation, and will improve the engineering of such systems.
\end{abstract}

\maketitle
\tableofcontents

\section{Introduction}
\vspace{-2mm}
Phase separation in elastic media is a ubiquitous phenomenon, which is relevant in synthetic systems to control micro-patterning~\cite{fernandez-rico2023Elastic,fernandez-rico2022Putting, style2018LiquidLiquid} and in biological cells, where droplets are embedded in the elastic cytoskeleton or chromatin~\cite{lee2022Mechanobiology,boddeker2022Nonspecific,lee2021Chromatin,lee2021Chromatin}.
While biological systems are typically dynamic and involve active processes, the simpler synthetic systems can exhibit regular stable structures.
These patterns harbor potential for metamaterials and structural color, particularly since they are easier to produce and manipulate than alternatives like self-assembly by block co-polymers~\cite{bates2012Multiblock} or chemical cross-linking~\cite{tran-cong1996ReactionInduced}.
In these applications, it is crucial to control the length scale, the quality, and the stability of the pattern.
Such control might be possible in a recent experiment, which found stable equilibrium patterns~\cite{fernandez-rico2023Elastic}.
However, the underlying mechanism for this elastic microphase separation is unclear, complicating further optimization.

The elastic microphase separation experiment~\cite{fernandez-rico2023Elastic} proceeds in two steps (\figref{M-fig:schematic_nonlocal_d}A):
First, a PDMS gel is soaked in oil at high temperatures for tens of hours until the system is equilibrated.
When the temperature is lowered in the second step, the sample develops bicontinuous structures, reminiscent of spinodal decomposition.
However, in contrast to spinodal decomposition, the length scale of the structure does not coarsen but stays arrested at roughly one to ten micrometers, depending on the gel's stiffness.
Interestingly, this transition is reversible and the pattern disappears upon reheating, suggesting a continuous phase transition.
Moreover, the resulting pattern is independent of the cooling rate, in contrast to other experiments on similar materials~\cite{rosowski2020Elastic,style2018LiquidLiquid}.
Consequently, the experiments should be explainable by an equilibrium theory that captures elastic deformations in PDMS due to oil droplets formed by phase separation.

In this paper, we propose a theoretical model that explains the experimental observations~\cite{fernandez-rico2023Elastic}.
The model combines the continuous density field of the elastic component, which naturally describe phase separation, with a nonlocal elasticity theory to capture the microstructure of the gel~\cite{kunin1982Elastic,kunin1983Elastic,eringen1992Vistas,eringen2004Nonlocal,maranganti2007Length}.
This approach allows us to capture the continuous phase transition to a patterned phase and predict its equilibrium period.

\section{Results}
To explain the experimental results~\cite{fernandez-rico2023Elastic} using an equilibrium theory, we define a free energy comprising contributions from elastic deformation as well as entropic and enthalpic contributions that can induce phase separation.
While elastic deformations are naturally described by the strain tensor $\boldsymbol{\epsilon}(\boldX)$ defined in a reference frame $\boldX$, phase separation is typically described by the volume fraction field $\phi(\boldsymbol{x})$ of the elastic component in the lab frame~$\boldsymbol{x}$.
For simplicity, we will focus on one-dimensional systems in this paper, where volume conservation connects the scalar strain $\epsilon$ to the fraction $\phi$ in the reference frame,
\begin{align}
    \label{M-eqn:strain_fraction_relation}
    \epsilon(X) & =\frac{\phi_0}{\phiE(X)} - 1
    \;,
\end{align}
where $\phi_0$ denotes the fraction in the relaxed homogeneous initial state.
The fraction $\phi(x)$ in the lab frame then follows from the coordinate transform $\dx/\dX=\epsilon(X)+1$.
This connection between strain~$\epsilon$ and volume fraction~$\phi$ permits a theory in terms of only one scalar field.

\subsection{Local elasticity models cannot explain equilibrium pattern}
We start by investigating a broad class of elastic models, where the elastic energy density is a function of strain~$\epsilon$.
Since $\epsilon$ can also be expressed in terms of the volume fraction~$\phiE$ (\Eqref{M-eqn:strain_fraction_relation}), the free energy of the system reads
\begin{align}
    F_\mathrm{local}[\phi] & = \frac{\kBT}{\volS} \int \Bigl[
        f_0(\phiE) + \kappa \, (\nabla \phiE)^2
        \Bigr] \dx
    \;,
    \label{M-eqn:fe_functional_generic}
\end{align}
where $k_\mathrm{B}$ is Boltzmann's constant, $T$ is the constant absolute temperature of the system, and $\volS$ is a relevant molecular volume, e.g., of the solvent molecules.
Here, $f_0$ captures the elastic energy density as well as molecular interactions and translational entropy associated with ordinary phase separation, while the second term proportional to the interfacial parameter $\kappa$ penalizes volume fraction gradients and gives rise to surface tension.
\Eqref{M-eqn:fe_functional_generic} is identical to basic models of phase separation without elastic contributions~\cite{cahn1958Free,weber2019Physics}.
Such models exhibit phase separation and subsequent coarsening to minimize interfacial costs, known as Ostwald ripening~\cite{voorhees1985Theory}.
While adding local elasticity alters $f_0(\phi)$, functions that minimize $F_\mathrm{local}$ can only have a single interface~\cite{carr1984Structured} and equilibrium patterns with finite length scales are thus impossible.
We show in the \supportingName{} \ref{si:local_cannot} that this result generalizes to higher dimensions.
Taken together, field theories based on local elasticity, including sophisticated non-linear finite strain models, cannot explain the equilibrium length scales observed in experiments.
\subsection{Microscopic picture suggests nonlocal elasticity theory}

\begin{figure}[t]
    \begin{center}
        \includegraphics[width = 1.0 \linewidth]{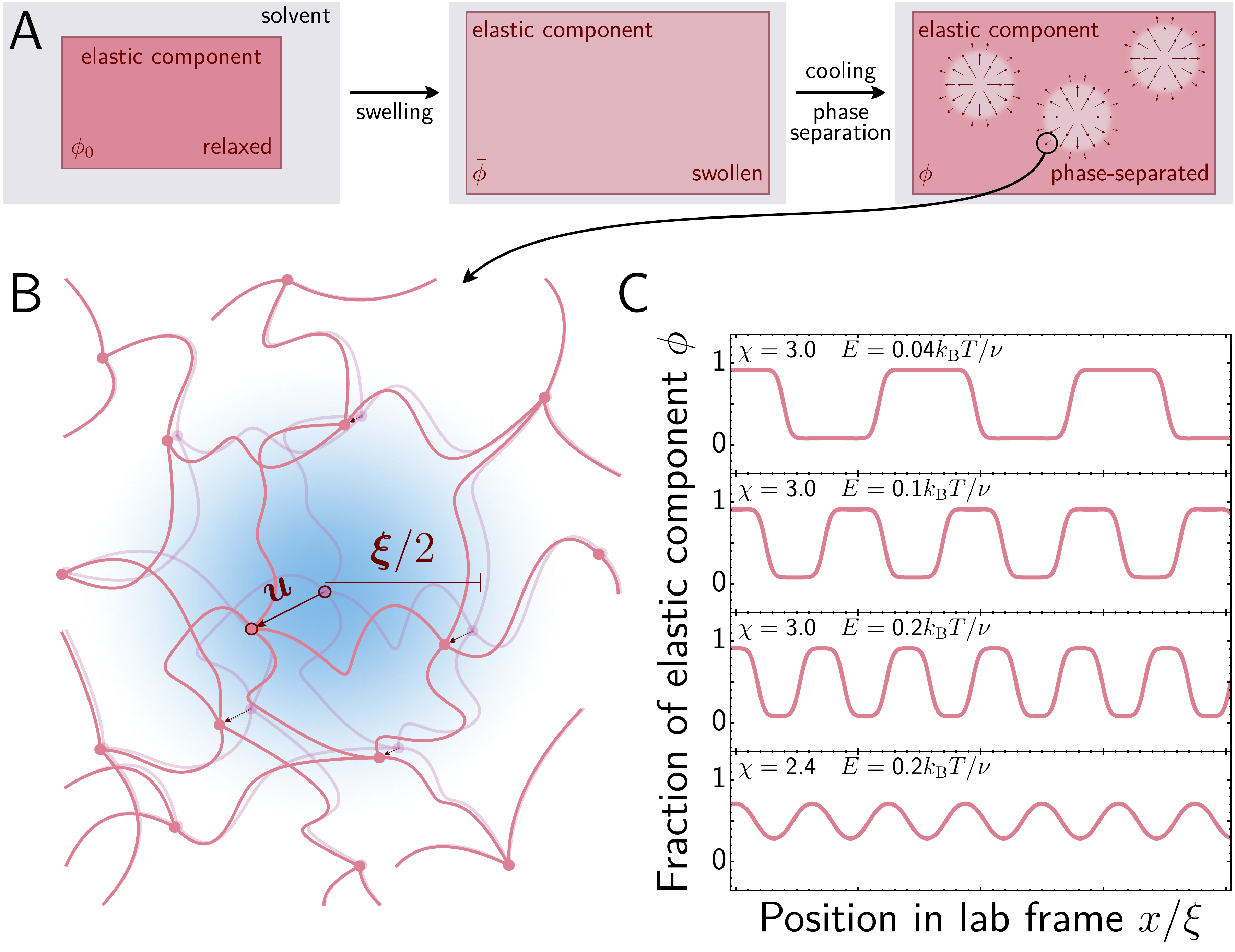}
    \end{center}
    \caption{
        \textbf{Nonlocal elasticity yields regular equilibrium patterns.}
        (A) Schematic picture of the experiment~\cite{fernandez-rico2023Elastic}:
        A relaxed elastic gel is swollen in a solvent at high temperature;
        After cooling, a regular pattern emerges.
        (B) Schematic of a polymer network displaying the displacement (opaque chains) from the reference state (transparent) after the central crosslink has been moved by $\boldu$.
        We model the forces transmitted along the network using a nonlocal convolution kernel (blue density) of size~$\xi$.
        (C) Equilibrium states for various stiffnesses $E$ and interaction parameters~$\chi$ for $\phi_0=1$, $\bar{\phi} = 0.5$, and $\kappa = 0.05\, \xi^{2}$.
    }
    \label{M-fig:schematic_nonlocal_d}
\end{figure}

Why do standard elastic theories fail to explain the observed patterns?
One answer is that only the interfacial parameters~$\kappa$ carries dimensions of length in \Eqref{M-eqn:fe_functional_generic}, so on dimensional grounds we cannot expect another length scale beyond the interfacial width to emerge.
While the interfacial width is typically governed by molecular sizes ($\sim\SI{1}{\nano\meter}$), realistic elastic meshes exhibit additional length scales like the mesh size  ($\sim\SI{10}{\nano\meter}$~\cite{maranganti2007Length,richbourg2020Swollen,ronceray2022Liquid}) and correlation lengths of spatial inhomogeneities ($\sim\SI{100}{\nano\meter}$~\cite{saalwachter2018Dynamicsbased,seiffert2017Origin,malodemolina2015Heterogeneity}).
Since the last two quantities are comparable to the pattern length scale (several $\SI{100}{\nano\meter}$ to several $\unit{\micro\meter}$~\cite{fernandez-rico2023Elastic}), we hypothesize that a characteristic length of the mesh is key for explaining the observed patterns.

If microscopic lengths of the elastic mesh are relevant, local elastic theories are insufficient~\cite{kunin1982Elastic,kunin1983Elastic,eringen1992Vistas,eringen2004Nonlocal,maranganti2007Length}.
This is because moving a particular crosslink transmits forces to connected crosslinks in the vicinity (see \figref{M-fig:schematic_nonlocal_d}B), implying stresses are no longer local, and the associated elastic energy cannot be expressed as a function of the strain.
Instead, the stress on a particular crosslink is now given by a sum over all connected crosslinks, whose contributions decay with distance~$X$ in the reference frame~\cite{kunin1982Elastic,eringen2004Nonlocal}.
In a continuous field theory, this nonlocal averaging is expressed as a convolution operation~\cite{kunin1982Elastic,eringen2004Nonlocal}.
Using a simple linear elastic model for the local stress $E \epsilon$, with elastic modulus $E$, we obtain the \emph{nonlocal stress}
\begin{align}
    \sigma_\mathrm{nonlocal}(X) & = E \int \epsilon(X') \, g_\xi(X'-X) \, \dXp
    \;,
    \label{M-eqn:nonlocal_strain}
\end{align}
where we choose a Gaussian convolution kernel~\cite{eringen2004Nonlocal,gopalakrishnan2013Wave},
\begin{align}
    g_\xi(X) & = \sqrt{\frac{2}{\pi \xi^2}}\exp\left(-\frac{2 X^2}{\xi^2}\right)
    \;,
    \label{M-eqn:kernel}
\end{align}
with a characteristic length~$\xi$, which quantifies the microscopic length of the gel~\cite{kunin1982Elastic,eringen2004Nonlocal,gopalakrishnan2013Wave}.
This nonlocal model can also be derived more rigorously, either generically (see \supportingName{} \ref{si:nonlocal}) or from a more explicit microscopic model~\cite{kunin1982Elastic,wu1993Improved}.
Note that the convolution is performed in the reference frame since the topology of the network, governing which crosslinks interact with each other, is determined in this unperturbed state.

The elastic energy density of the system is then given by the product of strain and nonlocal stress, so the free energy of the entire system reads
\begin{align}
    F_\mathrm{nonlocal}[\phi] & = F_\mathrm{local}[\phi]
    + \frac{1}{2} \int \!\epsilon(X) \, \sigma_\mathrm{nonlocal}(X) \, \dX
    \;,
    \label{M-eqn:f_elastic}
\end{align}
where $F_\mathrm{local}$ now only captures the contributions associated with phase separation.
To capture the essence of phase separation, we consider a simple Flory-Huggins model for the local free energy density~\cite{flory1942Thermodynamics,huggins1941Solutions,flory1950Statistical},
\begin{align}
    f_0(\phi) & = \phiE\log\phiE + (1-\phiE) \log (1-\phiE) + \chi \phiE (1-\phiE)
    \;,
    \label{M-eqn:f_FH}
\end{align}
where $1-\phiE$ is the solvent fraction.
Here, the first two terms capture entropic contributions, while the last term describes the interaction between elastic and solvent components, quantified by the Flory-Huggins parameter~$\chi$.
Taken together, \Eqsref{M-eqn:strain_fraction_relation}--\ref{M-eqn:f_FH} define the free energy $F_\mathrm{nonlocal}$ as a functional of the fraction~$\phi$ of the elastic component.

\subsection{Nonlocal elasticity enables equilibrium patterns}

We start by analyzing equilibrium states of the model by determining profiles $\phi(x)$ that minimize $F_\mathrm{nonlocal}$ using a numerical scheme described in the \supportingName{} \ref{si:numerics}.
Beside typical macroscopic phase separation, we also find periodic patterns for some parameter sets; see \figref{M-fig:schematic_nonlocal_d}C and \figref{S-fig:free_energy_curve}.
In soft systems (small stiffness~$E$), dilute regions, corresponding to solvent droplets, alternate with dense regions, where the elastic mesh is hardly strained ($\epsilon\ll1$).
In contrast, a harmonic profile emerges for stiff systems (large~$E$).
Taken together, the nonlocal elastic theory supports periodic patterns that qualitatively resemble the experimentally observed ones~\cite{fernandez-rico2023Elastic}.

\begin{figure}[t]
    \begin{center}
        \includegraphics[width = 1.0 \linewidth]{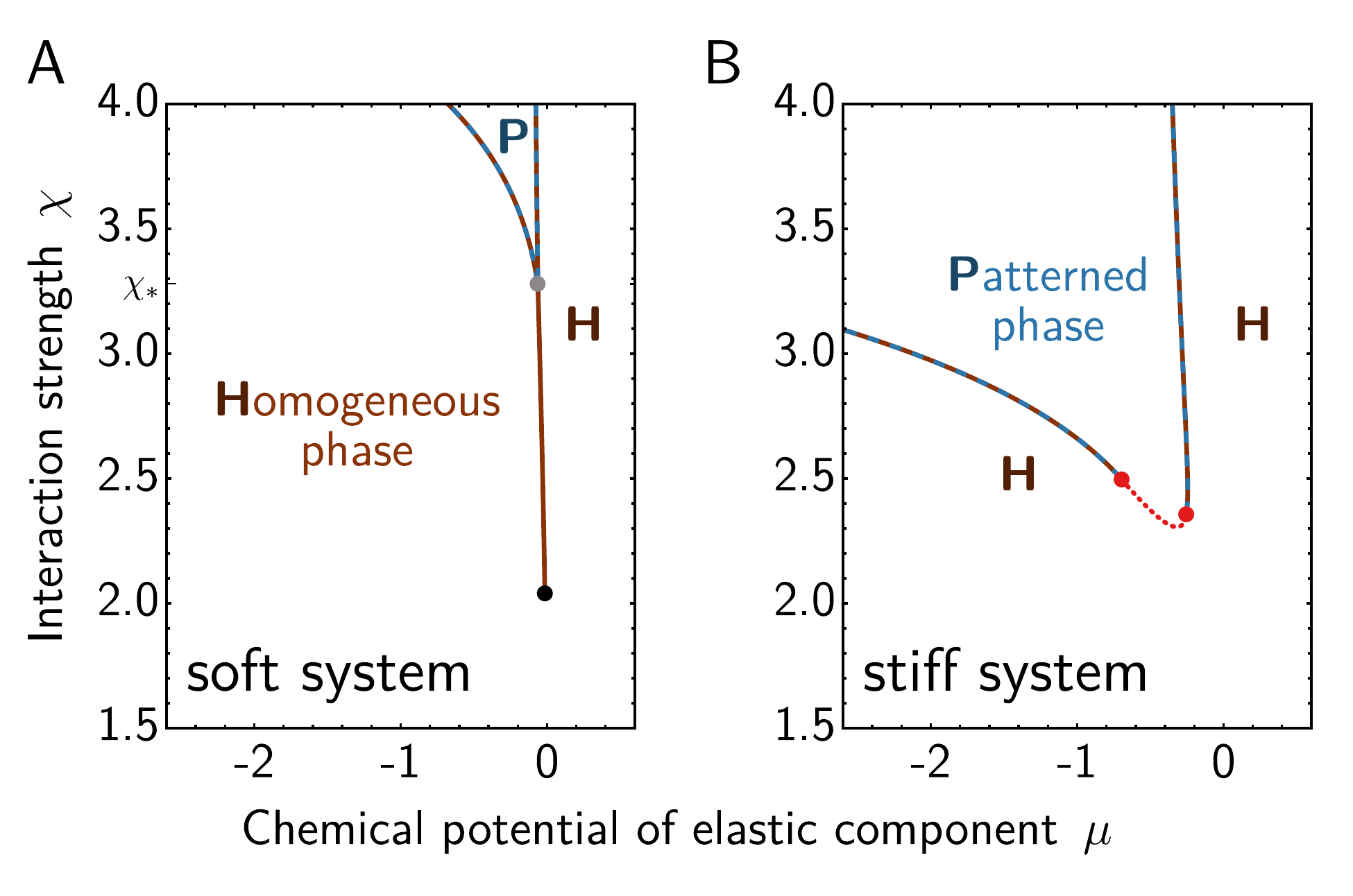}
    \end{center}
    \caption{
        \textbf{Grand-canonical phase diagrams reveal patterned phase.}
        (A) Phase diagram as a function of the chemical potential~$\mu$ and the interaction strength $\chi$ for $E=0.01 \, \kBT/\nu$.
        Homogeneous phases (region H) coexist on the brown line between the critical point of phase separation (black disk) and the triple point (gray disk), while the patterned phase (region P) coexists with the homogeneous phase on the blue-brown-dashed line.
        (B) Phase diagram as a function of $\mu$ and $\chi$ for $E=0.2 \, \kBT/\nu$.
        The binodal line separating the homogeneous and patterned phase exhibits either a first-order transition (blue-brown-dashed line) or a continuous transition (red dotted line with associated critical points marked by red disks; details in the \supportingName{} \ref{si:continuous}).
        (A--B) Model parameters are $\phi_0=1$ and $\kappa=0.05 \, \xi^{2}$.
    }
    \label{M-fig:phase_diagram_grandcanonical}
\end{figure}

To understand when periodic patterns form, we next investigate the simple case where components can freely exchange with a surrounding reservoir kept at fixed exchange chemical potential~$\mu$.
This situation allows solvent molecules to rush in and out of the system, adjusting the average fraction~$\bar{\phi}$ of the elastic component.
\figref{M-fig:phase_diagram_grandcanonical} shows two phase diagrams of this grand-canonical ensemble at different stiffnesses~$E$.
In the soft system (left panel), the phase diagram mostly resembles that of ordinary phase separation:
For weak interactions ($\chi < 2$), we find only a homogeneous phase and $\mu$ simply controls $\bar\phi$.
In contrast, above the critical point at $\chi\approx 2$ (black disk), we observe a first-order phase transition (brown line) between a dilute phase ($\mu \lesssim 0$) and a dense phase ($\mu \gtrsim 0$).
However, at even stronger interactions ($\chi \gtrsim 3.3$), an additional patterned phase (denoted by P) emerges, where the periodic patterns exhibit the lowest free energy.
The line of the first-order phase transitions between the patterned phase and the dilute or dense homogeneous phase (blue-brown-dashed curves) meet the line of the phase transition between the two homogeneous states at the triple point (gray disk), where these three states coexist.

\begin{figure*}
    \begin{center}
        \includegraphics[width = 1.0 \linewidth]{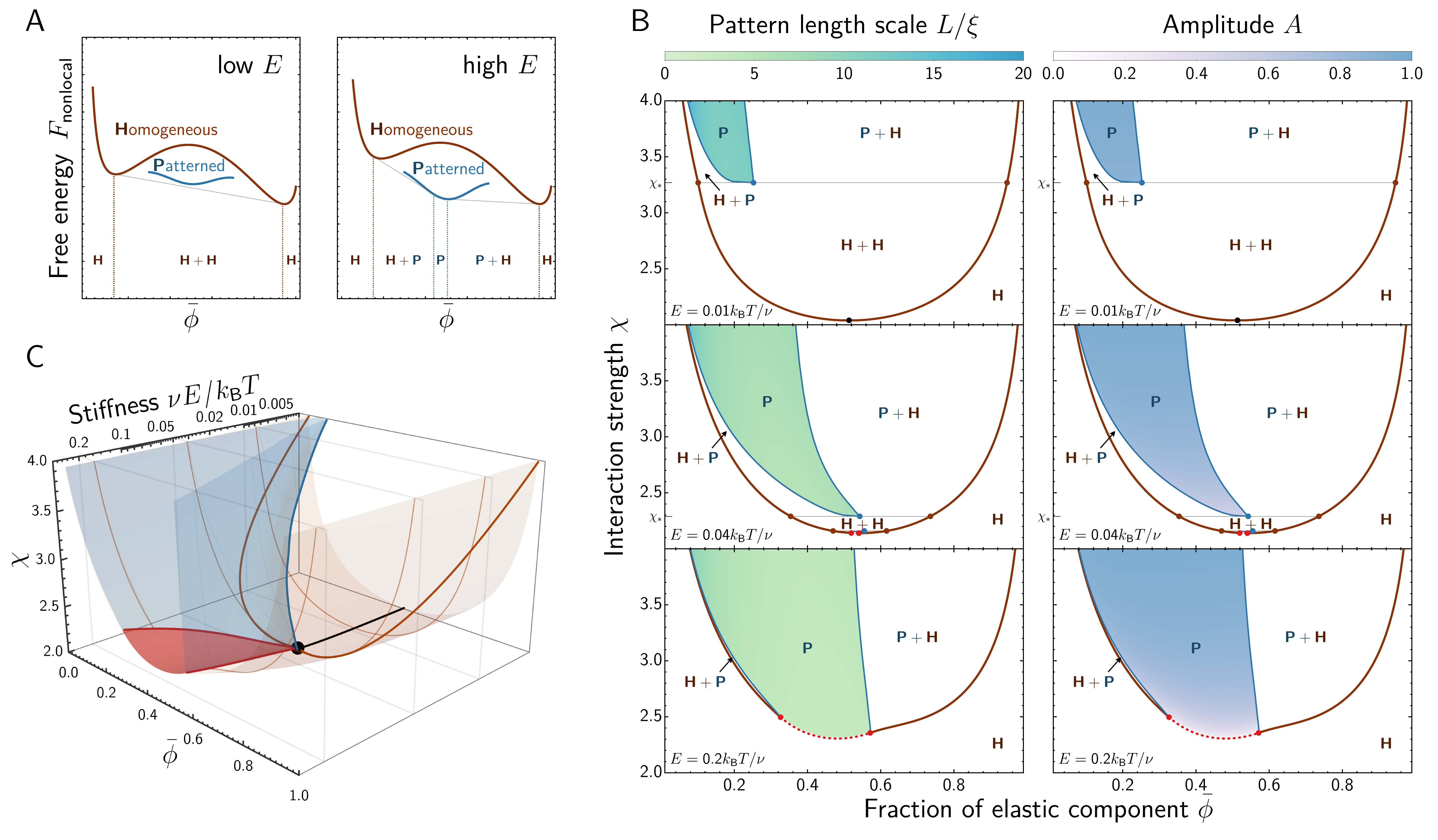}
    \end{center}
    \caption{
        \textbf{Closed systems exhibit phase coexistence.}
        (A)~Schematic free energy of homogeneous and patterned phases with common-tangent construction (thin gray lines) for two stiffnesses~$E$. \figref{S-fig:phase_diagram_1D_real} shows corresponding numerical results.
        (B)~Phase diagram as a function of the average fraction $\bar\phi$ of the elastic component and interaction strength $\chi$ for various $E$.
        Only the homogeneous phase (region H) is stable outside the binodal (brown line; black disk marks critical point) with a continuous phase transition at the red dotted part.
        Only the patterned phase (region P) is stable inside the blue lines with color codes indicating length scale and amplitude in the left and right column, respectively.
        Two indicated phases (H+P, P+H, H+H) coexist in other regions.
        The triple point corresponds to the tie line (thin gray line), where fractions $\bar{\phi}$ of coexisting homogeneous and patterned phases are marked by brown and blue disks, respectively.        (C)~Phase diagram as a function of $\bar\phi$, $\chi$, and $E$.
        The binodal of the homogeneous phase (brown surface) and the patterned phase (blue surface) overlap in the continuous phase transition (red surface).
        The critical points in panel B now correspond to critical lines, which all merge in the tricritical point (large black disk).
        A rotating version of the diagram is available as a \href{https://www.youtube.com/watch?v=cwOr_WsSISw}{movie}.
        (A--C) Model parameters are $\phi_0=1$ and $\kappa=0.05 \, \xi^{2}$.
    }
    \label{M-fig:phase_diagram}
\end{figure*}

The grand-canonical phase diagram of soft systems (left panel of \figref{M-fig:phase_diagram_grandcanonical}) qualitatively resembles simple pressure-temperature phase diagrams, e.g., of water.
Assuming that the chemical potential~$\mu$ plays the role of pressure and that the interaction~$\chi$ is negatively correlated with temperature, the dilute and dense homogeneous phases respectively correspond to the gas and liquid phases.
They become indistinguishable at the critical point at low interaction strength (corresponding to high temperatures).
In contrast, the patterned phase, with its periodic microstructure, resembles the solid phase.

The general form of the grand-canonical phase diagram persists for stiff system (right panel of \figref{M-fig:phase_diagram_grandcanonical}), although the parameter region of the patterned phase is much larger.
However, the first-order transition between the dilute and dense homogeneous phases disappears together with the normal critical point of phase separation.
Instead, we now find a continuous phase transition (dotted red line) between the homogeneous and the patterned phases, which we will discuss in more detail below.
Taken together, these phase diagrams suggest that stable patterned phases emerge for sufficiently large stiffness~$E$ and interaction~$\chi$ for intermediated~$\bar\phi$.

The grand-canonical ensemble that we discussed so far is suitable when the time scale of an experiment is long compared to the time scale of particle exchange with the reservoir.
In the experiments~\cite{fernandez-rico2023Elastic}, the initial swelling takes place over tens of hours with a measurable increase in size and mass, indicating that solvent soaks the sample until it is equilibrated with the surrounding bath.
In contrast, the temperature quench, during which the patterned phase is observed, takes place on a time scale of minutes without the solvent bath.
This suggests that this process is better described by a closed system.

\subsection{\PatNameAbbr{}ed and \homoName{}s coexist in closed systems}

In the closed system, corresponding to a canonical ensemble, the average fraction~$\bar\phi$ of elastic components, and thus also the average fraction of solvent, is fixed.
In this situation, we find that multiple different phases can coexist in the same system; see \figref{M-fig:phase_diagram}.
This is again reminiscent of phase separation, where the common-tangent construction reveals the fractions in coexisting homogeneous states.
Indeed, we find exactly this behavior in soft systems (left panel of \figref{M-fig:phase_diagram}A), where a dilute and dense phase coexist for fractions between the two vertical dotted lines, while the free energy of the patterned phase (blue line) is always larger and thus unfavorable.
The picture changes for larger stiffness (right panel of \figref{M-fig:phase_diagram}A), where the patterned phase has lower energy and we can construct two separate common tangents, which respectively connect the dilute and dense homogeneous phase with the patterned phase.
Analogously to phase separation, we thus expect situations in which a patterned phase coexists with a homogeneous phase (when $\bar{\phi}$ is in the region marked with H+P or P+H).
\figref{M-fig:phase_diagram}B corroborates this picture and shows various coexisting phases as a function of the stiffness~$E$ and the interaction strength~$\chi$.
Taken together, the main additional feature of the canonical phase diagrams is the coexistence of multiple phases, which was only possible exactly at the phase transition in the grand-canonical phase diagram.
\subsection{Higher stiffness and interaction strength stabilize \patName{} }

The canonical phase diagrams shown in \figref{M-fig:phase_diagram}B are complex, but they generally preserve three crucial aspects of the grand-canonical phase diagram shown in \figref{M-fig:phase_diagram_grandcanonical}: Higher stiffness (i) slightly favors the homogeneous phases, (ii) greatly expands the parameter region of the patterned phase, and (iii) induces a continuous phase transition.
The first point is illustrated by the binodal line of the homogeneous phase (thick brown lines and red dotted lines), which moves up with increasing stiffness~$E$, implying that larger interaction strengths~$\chi$ are necessary to stabilize inhomogeneous systems.
Inside the binodal line the system exhibits various behaviors, which can be categorized by $\chi$.
At a critical value $\chi_*$, the patterned phase (blue disk) coexists with the dilute and dense homogeneous phase (brown disks), and the associated tie line corresponds to the triple point in \figref{M-fig:phase_diagram_grandcanonical}.
For weaker interactions ($\chi<\chi_*$), we mostly observe coexistence of a dilute and dense homogeneous phase (region H+H), which corresponds to normal phase separation.
For stronger interactions ($\chi>\chi_*$), the system exhibits the patterned phase, either exclusively (colored region) or in coexistence with a homogeneous phase (regions H+P and P+H).
Larger stiffness~$E$ lowers the critical value $\chi_*$, thus expanding the parameter region where the patterned phase exists.
Eventually, for sufficiently large $E$, $\chi_*$ approaches the critical point of the binodal (gray point), a tiny region with patterned phase appears, and part of the binodal line becomes a continuous phase transition (red dotted line), reproducing the behavior predicted by the grand-canonical phase diagram of stiff systems (right panel in \figref{M-fig:phase_diagram_grandcanonical}).

The influence of stiffness~$E$ and interaction strength~$\chi$ becomes even more apparent in the three-dimensional phase diagram shown in \figref{M-fig:phase_diagram}C:
With increasing $E$, the $\chi$ associated with the critical point of phase separation (black line) increases slightly, whereas the states of three-phase coexistence (blue line and brown lines) shift to lower $\chi$.
All lines meet at the tricritical point (black sphere) for $E\approx 0.037 \, \kBT/\volS$, $\bar{\phi}\approx0.54$, and $\chi\approx 2.14$.
Increasing $E$ further, a part of the binodal line exhibits a continuous phase transition, which expands with larger~$E$.
The phase diagram thus summarizes three main aspects of our model:
First, the binodal line of phase separation, which is only weakly affected by $E$, determines whether the system can exhibit non-homogeneous states.
Second, if the system can be inhomogeneous, the stiffness~$E$ determines at what value of $\chi$ patterned phases emerge.
Third, for sufficiently large $E$, these patterned phases form immediately due to the continuous phase transition.

\subsection{Continuous phase transition explains experimental measurements}

\begin{figure}[t]
    \begin{center}
        \includegraphics[width = 1.0 \linewidth]{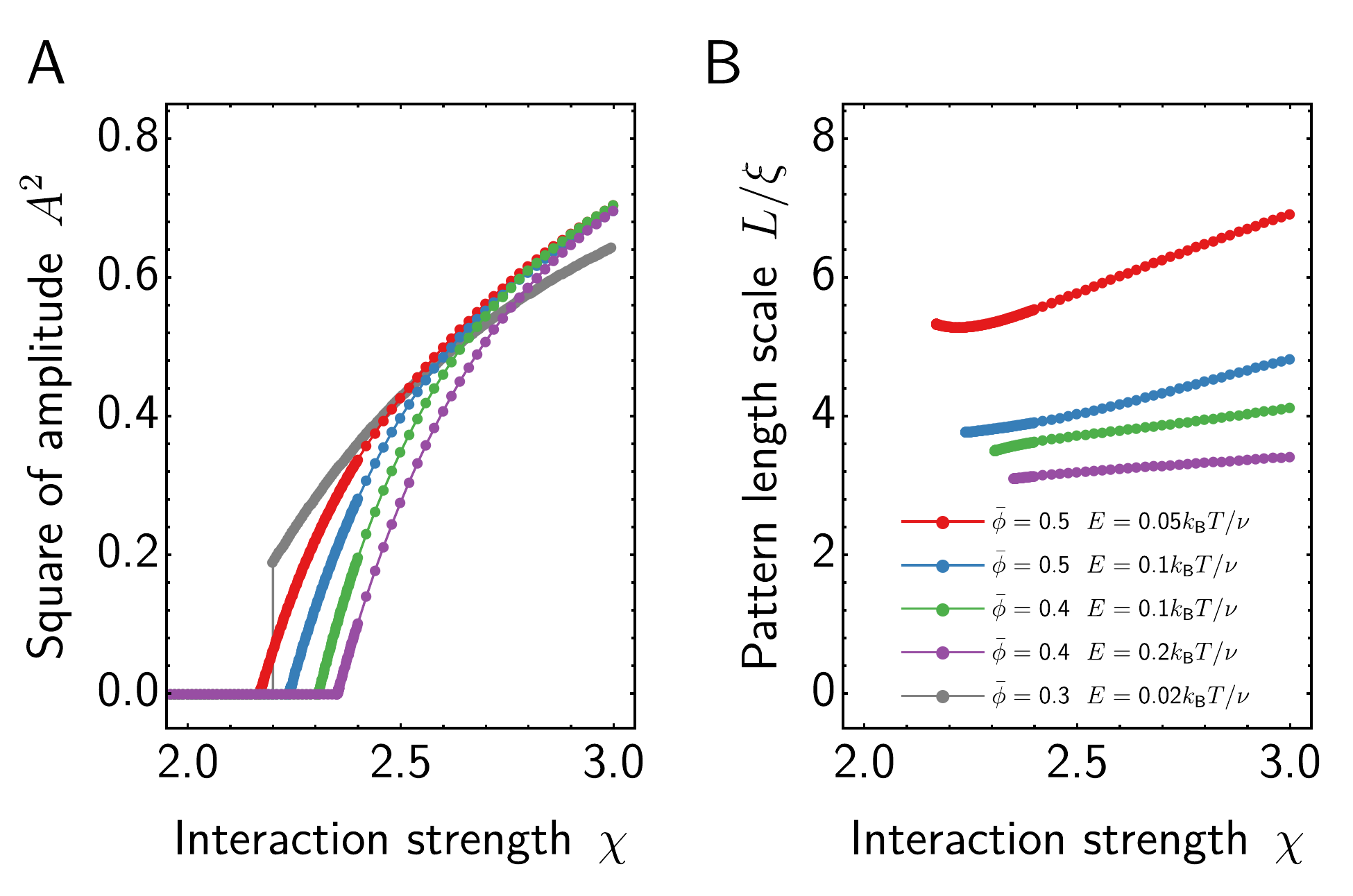}
    \end{center}
    \caption{
        \textbf{Continuous phase transition recovers experimental measurements.}
        Squared amplitude (panel A) and length scale (panel B) of periodic patterns as a function of interaction strength~$\chi$ for various parameters indicated in panel B, $\phi_0=1$, and $\kappa=0.05 \, \xi^{2}$.
        The amplitude indicates a continuous (colored data) and first-order (gray data) transition.
    }
    \label{M-fig:continuous_transition}
\end{figure}

The continuous phase transition that we identified at sufficiently large stiffness~$E$ implies that the system can change continuously from a homogeneous phase to a patterned phase when the interaction strength~$\chi$ is increased (corresponding to cooling).
Indeed, the amplitude of the predicted pattern vanishes near the transition (right panel of \figref{M-fig:phase_diagram}B), while the length scale stays finite (left panel of \figref{M-fig:phase_diagram}B).
This behavior is not expected for phase separating systems, where first-order transitions are typically, which are associated with a jump in observables (see gray dots in \figref{M-fig:continuous_transition}A for an example).

The continuous phase transition was already hypothesized for the experiments~\cite{fernandez-rico2023Elastic}, based on a lack of hysteresis and a continuous change of the contrast measured by light intensity.
To connect to experiments, we mimic the contrast using the square of the amplitude of the optimal volume fraction profile.
\figref{M-fig:continuous_transition}A and the right panel of \figref{M-fig:phase_diagram}B show that the contrast changes continuously from zero when the interaction strength~$\chi$ is increased for sufficiently stiff systems.
Moreover, \figref{M-fig:continuous_transition}B shows that the associated pattern length scale changes only slightly, consistent with the experiments.
Note that deviations in the form of the curves could stem from thermal fluctuations, finite resolution in the experiment, and also deviations in model details.

\subsection{Stiffness and interfacial cost control \patNameAbbr{} length scale }

\begin{figure}
    \begin{center}
        \includegraphics[width = 1.0 \linewidth]{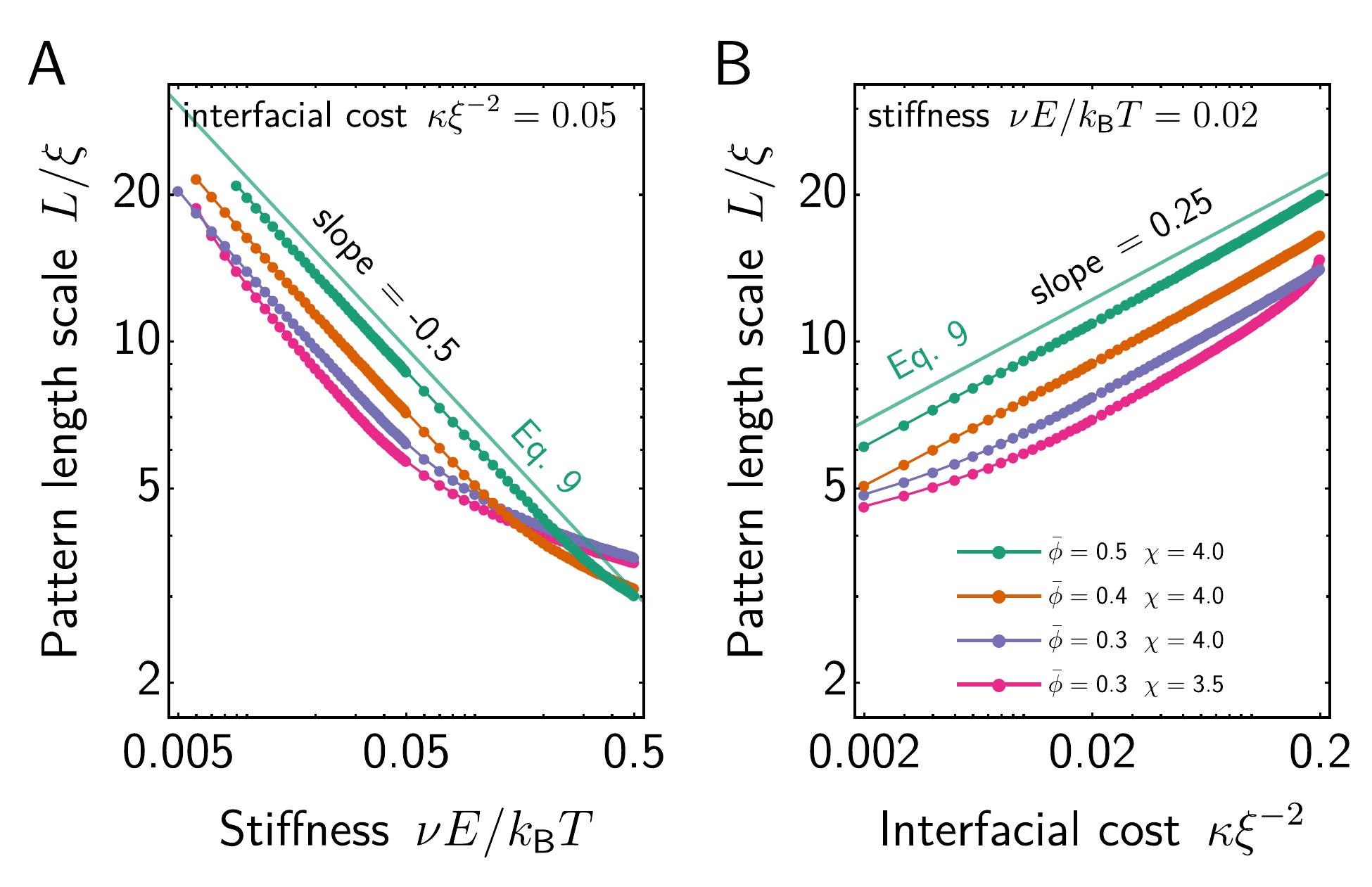}
    \end{center}
    \caption{
        \textbf{Pattern length scale exhibits scaling laws.}
        Length scale~$L$ as a function of stiffness $E$ (panel A) and interfacial parameter~$\kappa$ (panel B) for various parameters.
        Putative scaling laws are indicated and the prediction by \Eqref{M-eqn:length_scale} is shown for $\phi_0=1$, $\bar{\phi}=0.5$, $\chi=4$, and $\gamma \approx\kBT\kappa^{1/2} / \volS$ (green line).
    }
    \label{M-fig:length_scale}
\end{figure}

We next use the numerical minimization of the free energy~$F_\mathrm{nonlocal}$ to analyze how the length scale~$L$ of the patterned phase depends on parameters.
\figref{M-fig:length_scale} shows that $L$ decreases with larger stiffness~$E$ and increases with the interfacial cost parameterized by $\kappa$.
The data in \figref{M-fig:length_scale}A suggests the scaling $L/\xi \propto E^{-1/2}$ over a significant parameter range, which matches the experimental observations~\cite{fernandez-rico2023Elastic}.
Moreover, \figref{M-fig:length_scale}B suggests $L/\xi\propto  \xi^{-1/2} \kappa^{1/4}$, which has not been measured experimentally.
Taken together, the two scaling laws suggest that the equilibrium length scale emerges from a competition between elastic and interfacial energy.

The two scaling laws emerge qualitatively from a simple estimate of the elastic and interfacial energies:
Since shorter patterns have more interfaces, the interfacial energy per unit length is proportional to $\gamma L^{-1}$, with surface tension $\gamma \propto \kappa^{1/2}$~\cite{cahn1958Free}.
In contrast, the elastic energy of a single period originates from stretching a part of material from initial length~$\xi$ to final length $L$, resulting in an elastic energy density proportional to $EL\xi^{-1}$.
Minimizing the sum of these two energy densities with respect to $L$ results in $L/\xi\propto \xi^{-1/2} E^{-1/2}\kappa^{1/4}$, which explains the observed scalings qualitatively.

\subsection{Approximate model predicts length scale}

\begin{figure}
    \begin{center}
        \includegraphics[width = 1.0 \linewidth]{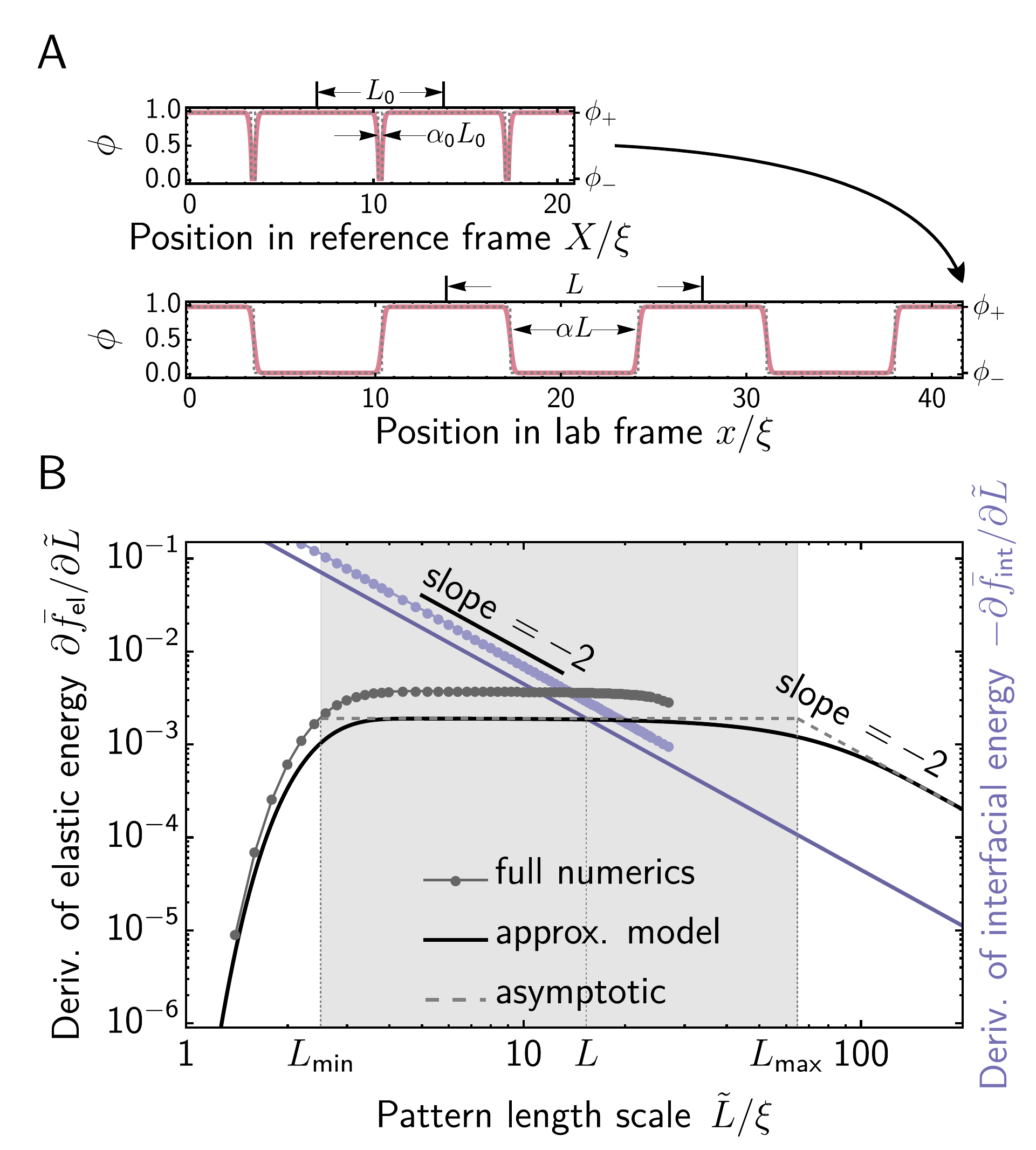}
    \end{center}
    \caption{
        \textbf{Approximate model explains scaling laws.}
        (A)
        Example for a volume fraction profile (pink lines) and the corresponding piecewise approximation (dotted gray lines) in the reference (top) and lab frame (bottom).
        (B)~Derivatives of the average energy density (in units of $\kBT\xi/\nu$) as a function of the pattern period~$\tilde L$.
        Shown are data from full numerics (dots), numerics for the piecewise profile (solid lines), and asymptotic functions (dashed lines) for the elastic (gray) and negative interfacial energy (violet).
        The stable length~$L$ corresponds to the crossing point of the elastic (black) and the interfacial terms (violet).
        Model parameters are $E=0.02\,\kBT/\nu $, $\kappa=0.05\, \xi^{2}$, $\phi_0=1$, $\bar\phiE=0.5$, and $\chi=4$.
    }
    \label{M-fig:box_function_approx}
\end{figure}

To understand the origin of the length scale~$L$ in more detail, we consider the limit of strong phase separation, where the interfacial width is small compared to~$L$; see \figref{M-fig:schematic_nonlocal_d}C.
We thus approximate the volume fraction profile~$\phi(x)$ of the elastic component by a periodic step function with fixed fractions $\phi_-$ and $\phi_+$; see dotted lines in \figref{M-fig:box_function_approx}A.
Material conservation implies that the relative size of these regions is dictated by the average fraction~$\bar\phi$ in the swollen state, so we can only vary the period~$\tilde L$ of the profile.
The stable period $L$ then corresponds to the $\tilde L$ that minimizes $F_\mathrm{nonlocal}$ given by \Eqref{M-eqn:f_elastic}, implying $F_\mathrm{nonlocal}'(L)=0$.
Since changing~$\tilde L$ does not affect the local free energy~$f_0$, we only investigate the average free energy of the interface, $\barfI(\tilde L) \approx 2\gamma \tilde{L}^{-1}$, and the average elastic free energy, $\barfE(\tilde L)= \frac{1}{2}{\tilde L}^{-1}  \int_0^{\tilde L_0} \sigma_{\mathrm{nonlocal}}(X) \epsilon(X) \dX$ where $\tilde L_0 = \frac{\bar{\phi}}{\phi_0}\tilde{L}$ is the period in the reference frame.
\figref{M-fig:box_function_approx}B shows the derivatives of these contributions with respect to $\tilde L$, indicating that they sum to zero for $\tilde L=L$.
We show in the \supportingName{} \ref{si:approx_model} that
\begin{align}
    \frac{\partial \barfE}{\partial \tilde L} & \approx \frac{E}{\xi}\cdot
    \begin{cases}
        0                                                                                                           & \tilde L<L_\mathrm{min}                \\[5pt]
        \frac{1}{\sqrt{2\pi}}\bigl(1 - \frac{\bar\phi}{\phi_+}\bigr)^2                                              & L_\mathrm{min}<\tilde L<L_\mathrm{max} \\[5pt]
        \frac{1}{\sqrt{8\pi}}\bigl(\frac{\phi_0}{\phi_{-}}-\frac{\phi_0}{\phi_{+}}\bigr)^2\frac{\xi^2}{\tilde{L}^2} & \tilde L>L_\mathrm{max}
    \end{cases} \;,
    \label{M-eqn:fe_derivatives_piecewise}
\end{align}
indicating three regimes bounded by
\begin{align}
    L_\mathrm{min} & = \sqrt{\frac{\pi}{2}}\,\frac{\phi_0}{\bar{\phi}}\,\xi                                                       & \text{and} &  &
    L_\mathrm{max} & = \sqrt{\frac{1}{2}} \, \frac{\phi_0}{\phi_{-}} \, \frac{\phi_{+}-\phi_{-}}{\phi_{+}-\bar{\phi}} \, \xi  \;.
    \label{M-eqn:length_bound}
\end{align}
\figref{M-fig:box_function_approx}B shows that this approximation of $\partial_{\tilde L} \barfE$ captures the main features of the full numerical data.

\figref{M-fig:box_function_approx}B suggests that stable patterns are mainly possible in the gray region ($L_\mathrm{min} < \tilde L < L_\mathrm{max}$), which we interpret further below.
In this region, we use \Eqref{M-eqn:fe_derivatives_piecewise} to solve $\partial_{\tilde L} \barfE + \partial_{\tilde L} \barfI =0$ for $\tilde L$, resulting in
\begin{align}
    L \approx (8\pi)^{\frac14}\frac{\phi_{+}}{\phi_{+}-\bar{\phi}} \, \left(\frac{\xi \gamma}{E}\right)^{\frac12}
    \;,
    \label{M-eqn:length_scale}
\end{align}
consistent with numerical results; see transparent green lines in \figref{M-fig:length_scale}.
This expression shows that the stable period $L$ is governed by the geometric mean of the elasto-capillary length $\gamma/E$ and the microscopic length~$\xi$.
Moreover, $L$ increases with a larger average fraction~$\bar\phi$ of the elastic component, i.e., less swelling.
In contrast, the fraction~$\phi_+$ has only a weak influence since it is close to $1$ in the case of strong phase separation, implying that the interaction strength~$\chi$ affects $L$ only weakly.

\subsection{Patterned phase is governed by reference state}
Finally, we use the approximate model to understand when the patterned phase emerges.
Here, it proves useful to interpret \Eqref{M-eqn:length_bound} in the reference frame, where the convolution of the nonlocal elastic energy takes place.
Defining the length $L_0=\frac{\bar{\phi}}{\phi_0}L$ in the reference frame and the associated fraction $\alpha_0=\frac{\phi_{-}}{\phi_0}(\phi_{+}-\bar{\phi})/(\phi_{+}-\phi_{-})$ occupied by the solvent droplet (\figref{M-fig:box_function_approx}A), we find
\begin{subequations}
    \label{M-eqn:length_bound_ref}
    \begin{align}
        L & >L_\mathrm{min} &  & \Leftrightarrow & L_0          & >\sqrt{\frac{\pi}{2}} \, \xi \qquad\text{and} \label{M-eqn:length_bound_ref_A} \\
        L & <L_\mathrm{max} &  & \Leftrightarrow & \alpha_0 L_0 & <\sqrt{\frac{1}{2}} \, \xi\;, \label{M-eqn:length_bound_ref_B}
    \end{align}
\end{subequations}
where the numerical pre-factors are very close to one.
The first condition ($L_0 \gtrsim \xi$) suggests that two solvent droplets need to be separated by more than $\xi$ in the reference frame since $L_0$ roughly estimates their separation; see \figref{M-fig:box_function_approx}A.
If droplets were closer, they would feel each other's deformations, which is apparently unfavorable.
In the extreme case ($L_0 \ll \xi$), the average elastic energy is almost constant, essentially because short-ranged variations are averaged by the comparatively large nonlocal kernel.
In contrast, the second condition implies that the droplet size in the reference frame ($\alpha_0L_0$) must be smaller than the microscopic length scale~$\xi$.
Assuming $\xi$ corresponds to the mesh correlation length, this suggests that the droplet can at most deform the correlated part of the mesh, which might correspond to large soft regions in natural meshes.
If droplets were larger ($\alpha_0L_0 \gg \xi$), nonlocal features would only be relevant at interfaces, so the system would behave as if it had only local elasticity and coarsen indefinitely.

This analysis highlights that the existence of the periodic pattern depends on the reference frame, while its length scale~$L$ also depends on the different stretch of the dilute and dense region; see \figref{M-fig:box_function_approx}A.
This observation suggest an intuitive explanation for the influence of the interaction~$\chi$:
Assuming that $\phi_-$ and $\phi_+$ correspond to equilibrium volume fractions and $\bar\phi=\frac12$ for simplicity, we find $\alpha_0 \propto \phi_-$, which decreases with larger~$\chi$.
Consequently, the lower bound $L_\mathrm{min}$ is unaffected, while $L_\mathrm{max}$ increases, consistent with our observation that the patterned phase forms easier at higher~$\chi$ and the scaling law given by \Eqref{M-eqn:length_scale} holds for broader parameter range with higher interaction strength (\figref{M-fig:length_scale}).

\section{Discussion}

We propose a theory that explains the experimentally observed elastic microphase separation~\cite{fernandez-rico2023Elastic} based on nonlocal elasticity, which captures aspects of the microscopic gel structure.
Within this theory, regular periodic patterns appear for sufficiently strong phase separation (large enough~$\chi$) and stiffness~$E$, while surface tension~$\gamma$ opposes the trend.
Essentially, solvent droplets inflate a region of the elastic mesh of the size of the microscopic length~$\xi$.
The pattern period~$L$ then results from a balance of elastic and interfacial energies, so that $L$ scales as the geometric mean between $\xi$ and the elasto-capillary length $\gamma/E$.
In contrast, the interaction strength~$\chi$, leading to phase separation in the first place, affects $L$ only weakly, but it determines whether the patterned phase is stable, similar to ordinary phase separation.
However, the normal first-order transition between the homogeneous and heterogeneous phase (at the binodal line) can now also exhibit a continuous phase transition.
Consequently, the patterned phase can appear with arbitrarily small amplitude in a reversible process.

Our model captures the main features of the experiment~\cite{fernandez-rico2023Elastic}, including the continuous phase transition leading to reversible dynamics.
Moreover, it explains that the pattern length scale~$L$ is independent of the cooling rate, only weakly affected by the final temperature, and decreases with stiffness~$E$.
Importantly, our model predicts that a structural length~$\xi$ of the mesh is essential for the emergence of the observed $L$.
Our numerics indicate that $L$ can be an order of magnitude larger than $\xi$, suggesting that $\xi$ could relate to observed correlation lengths of the order of a few hundred nanometers~\cite{saalwachter2018Dynamicsbased}.
Since $\xi$ is small compared to the distance between droplets (see \Eqref{M-eqn:length_bound_ref}), the nonlocal effects of elasticity do not affect droplet positioning.
Furthermore, we found that a coexisting homogeneous phase does not affect the free energy of the patterned phase strongly (see \supportingName{} \ref{si:numerics}), suggesting that the two phases can be interspersed, which would contribute to irregularity of the droplet placement in real systems.
In contrast, the observed variation in droplet size~\cite{fernandez-rico2023Elastic} likely originates from local heterogeneity in material properties, like $\xi$, $E$, and $\gamma$.
Taken together, our theory makes clear predictions that could be tested experimentally.

To capture the mesh's microstructure, we employ nonlocal elasticity~\cite{kunin1982Elastic,kunin1983Elastic,eringen1992Vistas,eringen2004Nonlocal,maranganti2007Length} based on a convolution of the stress field, which is similar to theories used in fracture mechanics~\cite{eringen1977CrackTip}.
Our work complements related theories, which either modeled pores explicitly~\cite{kothari2023Crucial,little2023Finitestraina,biswas2022Thermodynamics,ronceray2022Liquid,wei2020Modeling} or resorted to particle-based methods~\cite{curk2023Phase,zhang2021Mechanical}.
Nonlocality is generally responsible for the emergence of microstructure in multiple physical systems, such as the Ohta-Kawasaki model~\cite{ohta1986Equilibrium}, phase separation with electrostatic interaction~\cite{muratov2002Theory}, and also nonlocal elasticity~\cite{ren2000Finite,paulin2022Fluid}, e.g., to study polymeric materials~\cite{heyden2015Nonlocal,nikolov2007Origin}.
In contrast to the first two theories,
we use a convolution in the reference frame, capturing the microscopic topology of the elastic mesh.
More generally, the convolution kernel given by \Eqref{M-eqn:kernel} can be interpreted as a Green's function of a diffusion process in the reference frame, suggesting that the nonlocal elasticity is similar to the damage field introduced in fracture mechanics~\cite{bourdin2008Variational}.

We analyzed our model in the simple case of one dimension to highlight fundamental properties, but to capture experimental details, including various morphologies, we need to generalize the model to higher dimensions, which will require a tensorial convolution kernel~\cite{kunin1983Elastic}.
Moreover, we might require more realistic models of phase separation (including different molecular sizes and higher-order interactions terms) and elasticity (involving finite extensibility, viscoelasticity~\cite{tanaka2022Viscoelastic}, as well as plastic deformation, like fracture~\cite{kim2020Extreme,raayai-ardakani2019Intimate} and cavitation, which can lead to regular droplet patterns \cite{vidal-henriquez2021Cavitation}).
Finally, experimental systems exhibit heterogeneities in key model parameters including $\xi$, $E$, and $\gamma$, which will contribute to uncertainty and might even induce large scale rearrangements~\cite{rosowski2020Elastic,vidal-henriquez2020Theory}.
Such extended theories will allow us to compare the full pair correlation and scattering functions to experiments, shedding light on how we can manipulate this pattern forming system to control microstructures.

\tocless{\section*{Acknowledgments}}
We thank Carla Fern\'andez-Rico, Robert W. Style, and Eric R. Dufresne for helpful discussions.
We gratefully acknowledge funding from the Max Planck Society and the European Union (ERC, EmulSim, 101044662).

\bibliography{ps_em}

\begin{thebibliography}{50}%
\makeatletter
\providecommand \@ifxundefined [1]{%
 \@ifx{#1\undefined}
}%
\providecommand \@ifnum [1]{%
 \ifnum #1\expandafter \@firstoftwo
 \else \expandafter \@secondoftwo
 \fi
}%
\providecommand \@ifx [1]{%
 \ifx #1\expandafter \@firstoftwo
 \else \expandafter \@secondoftwo
 \fi
}%
\providecommand \natexlab [1]{#1}%
\providecommand \enquote  [1]{``#1''}%
\providecommand \bibnamefont  [1]{#1}%
\providecommand \bibfnamefont [1]{#1}%
\providecommand \citenamefont [1]{#1}%
\providecommand \href@noop [0]{\@secondoftwo}%
\providecommand \href [0]{\begingroup \@sanitize@url \@href}%
\providecommand \@href[1]{\@@startlink{#1}\@@href}%
\providecommand \@@href[1]{\endgroup#1\@@endlink}%
\providecommand \@sanitize@url [0]{\catcode `\\12\catcode `\$12\catcode
  `\&12\catcode `\#12\catcode `\^12\catcode `\_12\catcode `\%12\relax}%
\providecommand \@@startlink[1]{}%
\providecommand \@@endlink[0]{}%
\providecommand \url  [0]{\begingroup\@sanitize@url \@url }%
\providecommand \@url [1]{\endgroup\@href {#1}{\urlprefix }}%
\providecommand \urlprefix  [0]{URL }%
\providecommand \Eprint [0]{\href }%
\providecommand \doibase [0]{https://doi.org/}%
\providecommand \selectlanguage [0]{\@gobble}%
\providecommand \bibinfo  [0]{\@secondoftwo}%
\providecommand \bibfield  [0]{\@secondoftwo}%
\providecommand \translation [1]{[#1]}%
\providecommand \BibitemOpen [0]{}%
\providecommand \bibitemStop [0]{}%
\providecommand \bibitemNoStop [0]{.\EOS\space}%
\providecommand \EOS [0]{\spacefactor3000\relax}%
\providecommand \BibitemShut  [1]{\csname bibitem#1\endcsname}%
\let\auto@bib@innerbib\@empty
\bibitem [{\citenamefont {{Fern{\'a}ndez-Rico}}\ \emph
  {et~al.}(2023)\citenamefont {{Fern{\'a}ndez-Rico}}, \citenamefont
  {Schreiber}, \citenamefont {Oudich}, \citenamefont {Lorenz}, \citenamefont
  {Sicher}, \citenamefont {Sai}, \citenamefont {Heyden}, \citenamefont
  {Carrara}, \citenamefont {De~Lorenzis}, \citenamefont {Style},\ and\
  \citenamefont {Dufresne}}]{fernandez-rico2023Elastic}%
  \BibitemOpen
  \bibfield  {author} {\bibinfo {author} {\bibfnamefont {C.}~\bibnamefont
  {{Fern{\'a}ndez-Rico}}}, \bibinfo {author} {\bibfnamefont {S.}~\bibnamefont
  {Schreiber}}, \bibinfo {author} {\bibfnamefont {H.}~\bibnamefont {Oudich}},
  \bibinfo {author} {\bibfnamefont {C.}~\bibnamefont {Lorenz}}, \bibinfo
  {author} {\bibfnamefont {A.}~\bibnamefont {Sicher}}, \bibinfo {author}
  {\bibfnamefont {T.}~\bibnamefont {Sai}}, \bibinfo {author} {\bibfnamefont
  {S.}~\bibnamefont {Heyden}}, \bibinfo {author} {\bibfnamefont
  {P.}~\bibnamefont {Carrara}}, \bibinfo {author} {\bibfnamefont
  {L.}~\bibnamefont {De~Lorenzis}}, \bibinfo {author} {\bibfnamefont {R.~W.}\
  \bibnamefont {Style}},\ and\ \bibinfo {author} {\bibfnamefont {E.~R.}\
  \bibnamefont {Dufresne}},\ }\href {https://doi.org/10.48550/arXiv.2304.11419}
  {\bibinfo {title} {Elastic {{Microphase Separation Produces Robust
  Bicontinuous Materials}}}} (\bibinfo {year} {2023}),\ \Eprint
  {https://arxiv.org/abs/2304.11419} {arxiv:2304.11419 [cond-mat]} \BibitemShut
  {NoStop}%
\bibitem [{\citenamefont {{Fern{\'a}ndez-Rico}}\ \emph
  {et~al.}(2022)\citenamefont {{Fern{\'a}ndez-Rico}}, \citenamefont {Sai},
  \citenamefont {Sicher}, \citenamefont {Style},\ and\ \citenamefont
  {Dufresne}}]{fernandez-rico2022Putting}%
  \BibitemOpen
  \bibfield  {author} {\bibinfo {author} {\bibfnamefont {C.}~\bibnamefont
  {{Fern{\'a}ndez-Rico}}}, \bibinfo {author} {\bibfnamefont {T.}~\bibnamefont
  {Sai}}, \bibinfo {author} {\bibfnamefont {A.}~\bibnamefont {Sicher}},
  \bibinfo {author} {\bibfnamefont {R.~W.}\ \bibnamefont {Style}},\ and\
  \bibinfo {author} {\bibfnamefont {E.~R.}\ \bibnamefont {Dufresne}},\
  }\bibfield  {title} {\bibinfo {title} {Putting the {{Squeeze}} on {{Phase
  Separation}}},\ }\href {https://doi.org/10.1021/jacsau.1c00443} {\bibfield
  {journal} {\bibinfo  {journal} {JACS Au}\ }\textbf {\bibinfo {volume} {2}},\
  \bibinfo {pages} {66} (\bibinfo {year} {2022})}\BibitemShut {NoStop}%
\bibitem [{\citenamefont {Style}\ \emph {et~al.}(2018)\citenamefont {Style},
  \citenamefont {Sai}, \citenamefont {Fanelli}, \citenamefont {Ijavi},
  \citenamefont {{Smith-Mannschott}}, \citenamefont {Xu}, \citenamefont
  {Wilen},\ and\ \citenamefont {Dufresne}}]{style2018LiquidLiquid}%
  \BibitemOpen
  \bibfield  {author} {\bibinfo {author} {\bibfnamefont {R.~W.}\ \bibnamefont
  {Style}}, \bibinfo {author} {\bibfnamefont {T.}~\bibnamefont {Sai}}, \bibinfo
  {author} {\bibfnamefont {N.}~\bibnamefont {Fanelli}}, \bibinfo {author}
  {\bibfnamefont {M.}~\bibnamefont {Ijavi}}, \bibinfo {author} {\bibfnamefont
  {K.}~\bibnamefont {{Smith-Mannschott}}}, \bibinfo {author} {\bibfnamefont
  {Q.}~\bibnamefont {Xu}}, \bibinfo {author} {\bibfnamefont {L.~A.}\
  \bibnamefont {Wilen}},\ and\ \bibinfo {author} {\bibfnamefont {E.~R.}\
  \bibnamefont {Dufresne}},\ }\bibfield  {title} {\bibinfo {title}
  {Liquid-{{Liquid Phase Separation}} in an {{Elastic Network}}},\ }\href
  {https://doi.org/10.1103/PhysRevX.8.011028} {\bibfield  {journal} {\bibinfo
  {journal} {Phys. Rev. X}\ }\textbf {\bibinfo {volume} {8}},\ \bibinfo {pages}
  {011028} (\bibinfo {year} {2018})}\BibitemShut {NoStop}%
\bibitem [{\citenamefont {Lee}\ \emph {et~al.}(2022)\citenamefont {Lee},
  \citenamefont {Strom},\ and\ \citenamefont
  {Brangwynne}}]{lee2022Mechanobiology}%
  \BibitemOpen
  \bibfield  {author} {\bibinfo {author} {\bibfnamefont {D.~S.~W.}\
  \bibnamefont {Lee}}, \bibinfo {author} {\bibfnamefont {A.~R.}\ \bibnamefont
  {Strom}},\ and\ \bibinfo {author} {\bibfnamefont {C.~P.}\ \bibnamefont
  {Brangwynne}},\ }\bibfield  {title} {\bibinfo {title} {The mechanobiology of
  nuclear phase separation},\ }\href {https://doi.org/10.1063/5.0083286}
  {\bibfield  {journal} {\bibinfo  {journal} {APL Bioeng.}\ }\textbf {\bibinfo
  {volume} {6}},\ \bibinfo {pages} {021503} (\bibinfo {year}
  {2022})}\BibitemShut {NoStop}%
\bibitem [{\citenamefont {B{\"o}ddeker}\ \emph {et~al.}(2022)\citenamefont
  {B{\"o}ddeker}, \citenamefont {Rosowski}, \citenamefont {Berchtold},
  \citenamefont {Emmanouilidis}, \citenamefont {Han}, \citenamefont {Allain},
  \citenamefont {Style}, \citenamefont {Pelkmans},\ and\ \citenamefont
  {Dufresne}}]{boddeker2022Nonspecific}%
  \BibitemOpen
  \bibfield  {author} {\bibinfo {author} {\bibfnamefont {T.~J.}\ \bibnamefont
  {B{\"o}ddeker}}, \bibinfo {author} {\bibfnamefont {K.~A.}\ \bibnamefont
  {Rosowski}}, \bibinfo {author} {\bibfnamefont {D.}~\bibnamefont {Berchtold}},
  \bibinfo {author} {\bibfnamefont {L.}~\bibnamefont {Emmanouilidis}}, \bibinfo
  {author} {\bibfnamefont {Y.}~\bibnamefont {Han}}, \bibinfo {author}
  {\bibfnamefont {F.~H.~T.}\ \bibnamefont {Allain}}, \bibinfo {author}
  {\bibfnamefont {R.~W.}\ \bibnamefont {Style}}, \bibinfo {author}
  {\bibfnamefont {L.}~\bibnamefont {Pelkmans}},\ and\ \bibinfo {author}
  {\bibfnamefont {E.~R.}\ \bibnamefont {Dufresne}},\ }\bibfield  {title}
  {\bibinfo {title} {Non-specific adhesive forces between filaments and
  membraneless organelles},\ }\href
  {https://doi.org/10.1038/s41567-022-01537-8} {\bibfield  {journal} {\bibinfo
  {journal} {Nat. Phys.}\ }\textbf {\bibinfo {volume} {18}},\ \bibinfo {pages}
  {571} (\bibinfo {year} {2022})}\BibitemShut {NoStop}%
\bibitem [{\citenamefont {Lee}\ \emph {et~al.}(2021)\citenamefont {Lee},
  \citenamefont {Wingreen},\ and\ \citenamefont
  {Brangwynne}}]{lee2021Chromatin}%
  \BibitemOpen
  \bibfield  {author} {\bibinfo {author} {\bibfnamefont {D.~S.~W.}\
  \bibnamefont {Lee}}, \bibinfo {author} {\bibfnamefont {N.~S.}\ \bibnamefont
  {Wingreen}},\ and\ \bibinfo {author} {\bibfnamefont {C.~P.}\ \bibnamefont
  {Brangwynne}},\ }\bibfield  {title} {\bibinfo {title} {Chromatin mechanics
  dictates subdiffusion and coarsening dynamics of embedded condensates},\
  }\href {https://doi.org/10.1038/s41567-020-01125-8} {\bibfield  {journal}
  {\bibinfo  {journal} {Nat. Phys.}\ }\textbf {\bibinfo {volume} {17}},\
  \bibinfo {pages} {531} (\bibinfo {year} {2021})}\BibitemShut {NoStop}%
\bibitem [{\citenamefont {Bates}\ \emph {et~al.}(2012)\citenamefont {Bates},
  \citenamefont {Hillmyer}, \citenamefont {Lodge}, \citenamefont {Bates},
  \citenamefont {Delaney},\ and\ \citenamefont
  {Fredrickson}}]{bates2012Multiblock}%
  \BibitemOpen
  \bibfield  {author} {\bibinfo {author} {\bibfnamefont {F.~S.}\ \bibnamefont
  {Bates}}, \bibinfo {author} {\bibfnamefont {M.~A.}\ \bibnamefont {Hillmyer}},
  \bibinfo {author} {\bibfnamefont {T.~P.}\ \bibnamefont {Lodge}}, \bibinfo
  {author} {\bibfnamefont {C.~M.}\ \bibnamefont {Bates}}, \bibinfo {author}
  {\bibfnamefont {K.~T.}\ \bibnamefont {Delaney}},\ and\ \bibinfo {author}
  {\bibfnamefont {G.~H.}\ \bibnamefont {Fredrickson}},\ }\bibfield  {title}
  {\bibinfo {title} {Multiblock {{Polymers}}: {{Panacea}} or {{Pandora}}'s
  {{Box}}?},\ }\href {https://doi.org/10.1126/science.1215368} {\bibfield
  {journal} {\bibinfo  {journal} {Science}\ }\textbf {\bibinfo {volume}
  {336}},\ \bibinfo {pages} {434} (\bibinfo {year} {2012})}\BibitemShut
  {NoStop}%
\bibitem [{\citenamefont {{Tran-Cong}}\ and\ \citenamefont
  {Harada}(1996)}]{tran-cong1996ReactionInduced}%
  \BibitemOpen
  \bibfield  {author} {\bibinfo {author} {\bibfnamefont {Q.}~\bibnamefont
  {{Tran-Cong}}}\ and\ \bibinfo {author} {\bibfnamefont {A.}~\bibnamefont
  {Harada}},\ }\bibfield  {title} {\bibinfo {title} {Reaction-{{Induced
  Ordering Phenomena}} in {{Binary Polymer Mixtures}}},\ }\href
  {https://doi.org/10.1103/PhysRevLett.76.1162} {\bibfield  {journal} {\bibinfo
   {journal} {Phys. Rev. Lett.}\ }\textbf {\bibinfo {volume} {76}},\ \bibinfo
  {pages} {1162} (\bibinfo {year} {1996})}\BibitemShut {NoStop}%
\bibitem [{\citenamefont {Rosowski}\ \emph {et~al.}(2020)\citenamefont
  {Rosowski}, \citenamefont {Sai}, \citenamefont {{Vidal-Henriquez}},
  \citenamefont {Zwicker}, \citenamefont {Style},\ and\ \citenamefont
  {Dufresne}}]{rosowski2020Elastic}%
  \BibitemOpen
  \bibfield  {author} {\bibinfo {author} {\bibfnamefont {K.~A.}\ \bibnamefont
  {Rosowski}}, \bibinfo {author} {\bibfnamefont {T.}~\bibnamefont {Sai}},
  \bibinfo {author} {\bibfnamefont {E.}~\bibnamefont {{Vidal-Henriquez}}},
  \bibinfo {author} {\bibfnamefont {D.}~\bibnamefont {Zwicker}}, \bibinfo
  {author} {\bibfnamefont {R.~W.}\ \bibnamefont {Style}},\ and\ \bibinfo
  {author} {\bibfnamefont {E.~R.}\ \bibnamefont {Dufresne}},\ }\bibfield
  {title} {\bibinfo {title} {Elastic ripening and inhibition of
  liquid\textendash liquid phase separation},\ }\href
  {https://doi.org/10.1038/s41567-019-0767-2} {\bibfield  {journal} {\bibinfo
  {journal} {Nat. Phys.}\ }\textbf {\bibinfo {volume} {16}},\ \bibinfo {pages}
  {422} (\bibinfo {year} {2020})}\BibitemShut {NoStop}%
\bibitem [{\citenamefont {Kunin}(1982)}]{kunin1982Elastic}%
  \BibitemOpen
  \bibfield  {author} {\bibinfo {author} {\bibfnamefont {I.~A.}\ \bibnamefont
  {Kunin}},\ }\href {https://doi.org/10.1007/978-3-642-81748-9} {\emph
  {\bibinfo {title} {Elastic {{Media}} with {{Microstructure I}}}}},\ edited
  by\ \bibinfo {editor} {\bibfnamefont {E.}~\bibnamefont {Kr{\"o}ner}},
  \bibinfo {editor} {\bibfnamefont {M.}~\bibnamefont {Cardona}}, \bibinfo
  {editor} {\bibfnamefont {P.}~\bibnamefont {Fulde}},\ and\ \bibinfo {editor}
  {\bibfnamefont {H.-J.}\ \bibnamefont {Queisser}},\ \bibinfo {series}
  {Springer {{Series}} in {{Solid-State Sciences}}}, Vol.~\bibinfo {volume}
  {26}\ (\bibinfo  {publisher} {{Springer}},\ \bibinfo {address} {{Berlin,
  Heidelberg}},\ \bibinfo {year} {1982})\BibitemShut {NoStop}%
\bibitem [{\citenamefont {Kunin}(1983)}]{kunin1983Elastic}%
  \BibitemOpen
  \bibfield  {author} {\bibinfo {author} {\bibfnamefont {I.~A.}\ \bibnamefont
  {Kunin}},\ }\href {https://doi.org/10.1007/978-3-642-81960-5} {\emph
  {\bibinfo {title} {Elastic {{Media}} with {{Microstructure II}}}}},\ edited
  by\ \bibinfo {editor} {\bibfnamefont {E.}~\bibnamefont {Kr{\"o}ner}},
  \bibinfo {editor} {\bibfnamefont {M.}~\bibnamefont {Cardona}}, \bibinfo
  {editor} {\bibfnamefont {P.}~\bibnamefont {Fulde}},\ and\ \bibinfo {editor}
  {\bibfnamefont {H.-J.}\ \bibnamefont {Queisser}},\ \bibinfo {series}
  {Springer {{Series}} in {{Solid-State Sciences}}}, Vol.~\bibinfo {volume}
  {44}\ (\bibinfo  {publisher} {{Springer}},\ \bibinfo {address} {{Berlin,
  Heidelberg}},\ \bibinfo {year} {1983})\BibitemShut {NoStop}%
\bibitem [{\citenamefont {Eringen}(1992)}]{eringen1992Vistas}%
  \BibitemOpen
  \bibfield  {author} {\bibinfo {author} {\bibfnamefont {A.~C.}\ \bibnamefont
  {Eringen}},\ }\bibfield  {title} {\bibinfo {title} {Vistas of {{Nonlocal
  Continuum Physics}}},\ }\href {https://doi.org/10.1016/0020-7225(92)90165-D}
  {\bibfield  {journal} {\bibinfo  {journal} {Int. J. Engng Sci.}\ }\textbf
  {\bibinfo {volume} {30}},\ \bibinfo {pages} {1551} (\bibinfo {year}
  {1992})}\BibitemShut {NoStop}%
\bibitem [{\citenamefont {Eringen}(2004)}]{eringen2004Nonlocal}%
  \BibitemOpen
  \bibinfo {editor} {\bibfnamefont {A.~C.}\ \bibnamefont {Eringen}},\ ed.,\
  \href {https://doi.org/10.1007/b97697} {\emph {\bibinfo {title} {Nonlocal
  {{Continuum Field Theories}}}}}\ (\bibinfo  {publisher} {{Springer}},\
  \bibinfo {address} {{New York, NY}},\ \bibinfo {year} {2004})\BibitemShut
  {NoStop}%
\bibitem [{\citenamefont {Maranganti}\ and\ \citenamefont
  {Sharma}(2007)}]{maranganti2007Length}%
  \BibitemOpen
  \bibfield  {author} {\bibinfo {author} {\bibfnamefont {R.}~\bibnamefont
  {Maranganti}}\ and\ \bibinfo {author} {\bibfnamefont {P.}~\bibnamefont
  {Sharma}},\ }\bibfield  {title} {\bibinfo {title} {Length {{Scales}} at which
  {{Classical Elasticity Breaks Down}} for {{Various Materials}}},\ }\href
  {https://doi.org/10.1103/PhysRevLett.98.195504} {\bibfield  {journal}
  {\bibinfo  {journal} {Phys. Rev. Lett.}\ }\textbf {\bibinfo {volume} {98}},\
  \bibinfo {pages} {195504} (\bibinfo {year} {2007})}\BibitemShut {NoStop}%
\bibitem [{\citenamefont {Cahn}\ and\ \citenamefont
  {Hilliard}(1958)}]{cahn1958Free}%
  \BibitemOpen
  \bibfield  {author} {\bibinfo {author} {\bibfnamefont {J.~W.}\ \bibnamefont
  {Cahn}}\ and\ \bibinfo {author} {\bibfnamefont {J.~E.}\ \bibnamefont
  {Hilliard}},\ }\bibfield  {title} {\bibinfo {title} {Free {{Energy}} of a
  {{Nonuniform System}}. {{I}}. {{Interfacial Free Energy}}},\ }\href
  {https://doi.org/10.1063/1.1744102} {\bibfield  {journal} {\bibinfo
  {journal} {J. Chem. Phys.}\ }\textbf {\bibinfo {volume} {28}},\ \bibinfo
  {pages} {258} (\bibinfo {year} {1958})}\BibitemShut {NoStop}%
\bibitem [{\citenamefont {Weber}\ \emph {et~al.}(2019)\citenamefont {Weber},
  \citenamefont {Zwicker}, \citenamefont {J{\"u}licher},\ and\ \citenamefont
  {Lee}}]{weber2019Physics}%
  \BibitemOpen
  \bibfield  {author} {\bibinfo {author} {\bibfnamefont {C.~A.}\ \bibnamefont
  {Weber}}, \bibinfo {author} {\bibfnamefont {D.}~\bibnamefont {Zwicker}},
  \bibinfo {author} {\bibfnamefont {F.}~\bibnamefont {J{\"u}licher}},\ and\
  \bibinfo {author} {\bibfnamefont {C.~F.}\ \bibnamefont {Lee}},\ }\bibfield
  {title} {\bibinfo {title} {Physics of active emulsions},\ }\href
  {https://doi.org/10.1088/1361-6633/ab052b} {\bibfield  {journal} {\bibinfo
  {journal} {Rep. Prog. Phys.}\ }\textbf {\bibinfo {volume} {82}},\ \bibinfo
  {pages} {064601} (\bibinfo {year} {2019})}\BibitemShut {NoStop}%
\bibitem [{\citenamefont {Voorhees}(1985)}]{voorhees1985Theory}%
  \BibitemOpen
  \bibfield  {author} {\bibinfo {author} {\bibfnamefont {P.~W.}\ \bibnamefont
  {Voorhees}},\ }\bibfield  {title} {\bibinfo {title} {The {{Theory}} of
  {{Ostwald Ripening}}},\ }\href {https://doi.org/10.1007/BF01017860}
  {\bibfield  {journal} {\bibinfo  {journal} {J. Stat. Phys.}\ }\textbf
  {\bibinfo {volume} {38}},\ \bibinfo {pages} {231} (\bibinfo {year}
  {1985})}\BibitemShut {NoStop}%
\bibitem [{\citenamefont {Carr}\ \emph {et~al.}(1984)\citenamefont {Carr},
  \citenamefont {Gurtin},\ and\ \citenamefont {Slemrod}}]{carr1984Structured}%
  \BibitemOpen
  \bibfield  {author} {\bibinfo {author} {\bibfnamefont {J.}~\bibnamefont
  {Carr}}, \bibinfo {author} {\bibfnamefont {M.~E.}\ \bibnamefont {Gurtin}},\
  and\ \bibinfo {author} {\bibfnamefont {M.}~\bibnamefont {Slemrod}},\
  }\bibfield  {title} {\bibinfo {title} {Structured {{Phase Transitions}} on a
  {{Finite Interval}}},\ }\href {https://doi.org/10.1007/BF00280031} {\bibfield
   {journal} {\bibinfo  {journal} {Arch. Rational Mech. Anal.}\ }\textbf
  {\bibinfo {volume} {86}},\ \bibinfo {pages} {317} (\bibinfo {year}
  {1984})}\BibitemShut {NoStop}%
\bibitem [{\citenamefont {Richbourg}\ and\ \citenamefont
  {Peppas}(2020)}]{richbourg2020Swollen}%
  \BibitemOpen
  \bibfield  {author} {\bibinfo {author} {\bibfnamefont {N.~R.}\ \bibnamefont
  {Richbourg}}\ and\ \bibinfo {author} {\bibfnamefont {N.~A.}\ \bibnamefont
  {Peppas}},\ }\bibfield  {title} {\bibinfo {title} {The swollen polymer
  network hypothesis: {{Quantitative}} models of hydrogel swelling, stiffness,
  and solute transport},\ }\href
  {https://doi.org/10.1016/j.progpolymsci.2020.101243} {\bibfield  {journal}
  {\bibinfo  {journal} {Prog. Polym. Sci.}\ }\textbf {\bibinfo {volume}
  {105}},\ \bibinfo {pages} {101243} (\bibinfo {year} {2020})}\BibitemShut
  {NoStop}%
\bibitem [{\citenamefont {Ronceray}\ \emph {et~al.}(2022)\citenamefont
  {Ronceray}, \citenamefont {Mao}, \citenamefont {Ko{\v s}mrlj},\ and\
  \citenamefont {Haataja}}]{ronceray2022Liquid}%
  \BibitemOpen
  \bibfield  {author} {\bibinfo {author} {\bibfnamefont {P.}~\bibnamefont
  {Ronceray}}, \bibinfo {author} {\bibfnamefont {S.}~\bibnamefont {Mao}},
  \bibinfo {author} {\bibfnamefont {A.}~\bibnamefont {Ko{\v s}mrlj}},\ and\
  \bibinfo {author} {\bibfnamefont {M.~P.}\ \bibnamefont {Haataja}},\
  }\bibfield  {title} {\bibinfo {title} {Liquid demixing in elastic networks:
  {{Cavitation}}, permeation, or size selection?},\ }\href
  {https://doi.org/10.1209/0295-5075/ac56ac} {\bibfield  {journal} {\bibinfo
  {journal} {EPL}\ }\textbf {\bibinfo {volume} {137}},\ \bibinfo {pages}
  {67001} (\bibinfo {year} {2022})}\BibitemShut {NoStop}%
\bibitem [{\citenamefont {Saalw{\"a}chter}\ and\ \citenamefont
  {Seiffert}(2018)}]{saalwachter2018Dynamicsbased}%
  \BibitemOpen
  \bibfield  {author} {\bibinfo {author} {\bibfnamefont {K.}~\bibnamefont
  {Saalw{\"a}chter}}\ and\ \bibinfo {author} {\bibfnamefont {S.}~\bibnamefont
  {Seiffert}},\ }\bibfield  {title} {\bibinfo {title} {Dynamics-based
  assessment of nanoscopic polymer-network mesh structures and their defects},\
  }\href {https://doi.org/10.1039/C7SM02444D} {\bibfield  {journal} {\bibinfo
  {journal} {Soft Matter}\ }\textbf {\bibinfo {volume} {14}},\ \bibinfo {pages}
  {1976} (\bibinfo {year} {2018})}\BibitemShut {NoStop}%
\bibitem [{\citenamefont {Seiffert}(2017)}]{seiffert2017Origin}%
  \BibitemOpen
  \bibfield  {author} {\bibinfo {author} {\bibfnamefont {S.}~\bibnamefont
  {Seiffert}},\ }\bibfield  {title} {\bibinfo {title} {Origin of nanostructural
  inhomogeneity in polymer-network gels},\ }\href
  {https://doi.org/10.1039/C7PY01035D} {\bibfield  {journal} {\bibinfo
  {journal} {Polym. Chem.}\ }\textbf {\bibinfo {volume} {8}},\ \bibinfo {pages}
  {4472} (\bibinfo {year} {2017})}\BibitemShut {NoStop}%
\bibitem [{\citenamefont {{Malo de Molina}}\ \emph {et~al.}(2015)\citenamefont
  {{Malo de Molina}}, \citenamefont {Lad},\ and\ \citenamefont
  {Helgeson}}]{malodemolina2015Heterogeneity}%
  \BibitemOpen
  \bibfield  {author} {\bibinfo {author} {\bibfnamefont {P.}~\bibnamefont
  {{Malo de Molina}}}, \bibinfo {author} {\bibfnamefont {S.}~\bibnamefont
  {Lad}},\ and\ \bibinfo {author} {\bibfnamefont {M.~E.}\ \bibnamefont
  {Helgeson}},\ }\bibfield  {title} {\bibinfo {title} {Heterogeneity and its
  {{Influence}} on the {{Properties}} of {{Difunctional Poly}}(ethylene glycol)
  {{Hydrogels}}: {{Structure}} and {{Mechanics}}},\ }\href
  {https://doi.org/10.1021/acs.macromol.5b01115} {\bibfield  {journal}
  {\bibinfo  {journal} {Macromolecules}\ }\textbf {\bibinfo {volume} {48}},\
  \bibinfo {pages} {5402} (\bibinfo {year} {2015})}\BibitemShut {NoStop}%
\bibitem [{\citenamefont {Gopalakrishnan}\ and\ \citenamefont
  {Narendar}(2013)}]{gopalakrishnan2013Wave}%
  \BibitemOpen
  \bibfield  {author} {\bibinfo {author} {\bibfnamefont {S.}~\bibnamefont
  {Gopalakrishnan}}\ and\ \bibinfo {author} {\bibfnamefont {S.}~\bibnamefont
  {Narendar}},\ }\href {https://doi.org/10.1007/978-3-319-01032-8} {\emph
  {\bibinfo {title} {Wave {{Propagation}} in {{Nanostructures}}: {{Nonlocal
  Continuum Mechanics Formulations}}}}},\ {{NanoScience}} and {{Technology}}\
  (\bibinfo  {publisher} {{Springer International Publishing}},\ \bibinfo
  {address} {{Cham}},\ \bibinfo {year} {2013})\BibitemShut {NoStop}%
\bibitem [{\citenamefont {Wu}\ and\ \citenamefont {Van
  Der~Giessen}(1993)}]{wu1993Improved}%
  \BibitemOpen
  \bibfield  {author} {\bibinfo {author} {\bibfnamefont {P.~D.}\ \bibnamefont
  {Wu}}\ and\ \bibinfo {author} {\bibfnamefont {E.}~\bibnamefont {Van
  Der~Giessen}},\ }\bibfield  {title} {\bibinfo {title} {On {{Improved Network
  Models}} for {{Rubber Elasticity}} and {{Their Applications}} to
  {{Orientation Hardening}} in {{Glassy Polymers}}},\ }\href
  {https://doi.org/10.1016/0022-5096(93)90043-F} {\bibfield  {journal}
  {\bibinfo  {journal} {J. Mech. Phys. Solids}\ }\textbf {\bibinfo {volume}
  {41}},\ \bibinfo {pages} {427} (\bibinfo {year} {1993})}\BibitemShut
  {NoStop}%
\bibitem [{\citenamefont {Flory}(1942)}]{flory1942Thermodynamics}%
  \BibitemOpen
  \bibfield  {author} {\bibinfo {author} {\bibfnamefont {P.~J.}\ \bibnamefont
  {Flory}},\ }\bibfield  {title} {\bibinfo {title} {Thermodynamics of {{High
  Polymer Solutions}}},\ }\href {https://doi.org/10.1063/1.1723621} {\bibfield
  {journal} {\bibinfo  {journal} {J. Chem. Phys.}\ }\textbf {\bibinfo {volume}
  {10}},\ \bibinfo {pages} {51} (\bibinfo {year} {1942})}\BibitemShut {NoStop}%
\bibitem [{\citenamefont {Huggins}(1941)}]{huggins1941Solutions}%
  \BibitemOpen
  \bibfield  {author} {\bibinfo {author} {\bibfnamefont {M.~L.}\ \bibnamefont
  {Huggins}},\ }\bibfield  {title} {\bibinfo {title} {Solutions of {{Long Chain
  Compounds}}},\ }\href {https://doi.org/10.1063/1.1750930} {\bibfield
  {journal} {\bibinfo  {journal} {J. Chem. Phys.}\ }\textbf {\bibinfo {volume}
  {9}},\ \bibinfo {pages} {440} (\bibinfo {year} {1941})}\BibitemShut {NoStop}%
\bibitem [{\citenamefont {Flory}(1950)}]{flory1950Statistical}%
  \BibitemOpen
  \bibfield  {author} {\bibinfo {author} {\bibfnamefont {P.~J.}\ \bibnamefont
  {Flory}},\ }\bibfield  {title} {\bibinfo {title} {Statistical {{Mechanics}}
  of {{Swelling}} of {{Network Structures}}},\ }\href
  {https://doi.org/10.1063/1.1747424} {\bibfield  {journal} {\bibinfo
  {journal} {J. Chem. Phys.}\ }\textbf {\bibinfo {volume} {18}},\ \bibinfo
  {pages} {108} (\bibinfo {year} {1950})}\BibitemShut {NoStop}%
\bibitem [{\citenamefont {Eringen}\ \emph {et~al.}(1977)\citenamefont
  {Eringen}, \citenamefont {Speziale},\ and\ \citenamefont
  {Kim}}]{eringen1977CrackTip}%
  \BibitemOpen
  \bibfield  {author} {\bibinfo {author} {\bibfnamefont {A.~C.}\ \bibnamefont
  {Eringen}}, \bibinfo {author} {\bibfnamefont {C.~G.}\ \bibnamefont
  {Speziale}},\ and\ \bibinfo {author} {\bibfnamefont {B.~S.}\ \bibnamefont
  {Kim}},\ }\bibfield  {title} {\bibinfo {title} {Crack-{{Tip Problem}} in
  {{Non-Local Elasticity}}},\ }\href
  {https://doi.org/10.1016/0022-5096(77)90002-3} {\bibfield  {journal}
  {\bibinfo  {journal} {J. Mech. Phys. Solids}\ }\textbf {\bibinfo {volume}
  {25}},\ \bibinfo {pages} {339} (\bibinfo {year} {1977})}\BibitemShut
  {NoStop}%
\bibitem [{\citenamefont {Kothari}\ and\ \citenamefont
  {Cohen}(2023)}]{kothari2023Crucial}%
  \BibitemOpen
  \bibfield  {author} {\bibinfo {author} {\bibfnamefont {M.}~\bibnamefont
  {Kothari}}\ and\ \bibinfo {author} {\bibfnamefont {T.}~\bibnamefont
  {Cohen}},\ }\bibfield  {title} {\bibinfo {title} {The crucial role of
  elasticity in regulating liquid\textendash liquid phase separation in
  cells},\ }\href {https://doi.org/10.1007/s10237-022-01670-6} {\bibfield
  {journal} {\bibinfo  {journal} {Biomech. Model Mechanobiol.}\ }\textbf
  {\bibinfo {volume} {22}},\ \bibinfo {pages} {645} (\bibinfo {year}
  {2023})}\BibitemShut {NoStop}%
\bibitem [{\citenamefont {Little}\ \emph {et~al.}(2023)\citenamefont {Little},
  \citenamefont {Levine}, \citenamefont {Singh},\ and\ \citenamefont
  {Bruinsma}}]{little2023Finitestraina}%
  \BibitemOpen
  \bibfield  {author} {\bibinfo {author} {\bibfnamefont {J.}~\bibnamefont
  {Little}}, \bibinfo {author} {\bibfnamefont {A.~J.}\ \bibnamefont {Levine}},
  \bibinfo {author} {\bibfnamefont {A.~R.}\ \bibnamefont {Singh}},\ and\
  \bibinfo {author} {\bibfnamefont {R.}~\bibnamefont {Bruinsma}},\ }\bibfield
  {title} {\bibinfo {title} {Finite-strain elasticity theory and liquid-liquid
  phase separation in compressible gels},\ }\href
  {https://doi.org/10.1103/PhysRevE.107.024418} {\bibfield  {journal} {\bibinfo
   {journal} {Phys. Rev. E}\ }\textbf {\bibinfo {volume} {107}},\ \bibinfo
  {pages} {024418} (\bibinfo {year} {2023})}\BibitemShut {NoStop}%
\bibitem [{\citenamefont {Biswas}\ \emph {et~al.}(2022)\citenamefont {Biswas},
  \citenamefont {Mukherjee},\ and\ \citenamefont
  {Chakrabarti}}]{biswas2022Thermodynamics}%
  \BibitemOpen
  \bibfield  {author} {\bibinfo {author} {\bibfnamefont {S.}~\bibnamefont
  {Biswas}}, \bibinfo {author} {\bibfnamefont {B.}~\bibnamefont {Mukherjee}},\
  and\ \bibinfo {author} {\bibfnamefont {B.}~\bibnamefont {Chakrabarti}},\
  }\bibfield  {title} {\bibinfo {title} {Thermodynamics predicts a stable
  microdroplet phase in polymer\textendash gel mixtures undergoing elastic
  phase separation},\ }\href {https://doi.org/10.1039/D2SM01101H} {\bibfield
  {journal} {\bibinfo  {journal} {Soft Matter}\ }\textbf {\bibinfo {volume}
  {18}},\ \bibinfo {pages} {8117} (\bibinfo {year} {2022})}\BibitemShut
  {NoStop}%
\bibitem [{\citenamefont {Wei}\ \emph {et~al.}(2020)\citenamefont {Wei},
  \citenamefont {Zhou}, \citenamefont {Wang},\ and\ \citenamefont
  {Meng}}]{wei2020Modeling}%
  \BibitemOpen
  \bibfield  {author} {\bibinfo {author} {\bibfnamefont {X.}~\bibnamefont
  {Wei}}, \bibinfo {author} {\bibfnamefont {J.}~\bibnamefont {Zhou}}, \bibinfo
  {author} {\bibfnamefont {Y.}~\bibnamefont {Wang}},\ and\ \bibinfo {author}
  {\bibfnamefont {F.}~\bibnamefont {Meng}},\ }\bibfield  {title} {\bibinfo
  {title} {Modeling {{Elastically Mediated Liquid-Liquid Phase Separation}}},\
  }\href {https://doi.org/10.1103/PhysRevLett.125.268001} {\bibfield  {journal}
  {\bibinfo  {journal} {Phys. Rev. Lett.}\ }\textbf {\bibinfo {volume} {125}},\
  \bibinfo {pages} {268001} (\bibinfo {year} {2020})}\BibitemShut {NoStop}%
\bibitem [{\citenamefont {Curk}\ and\ \citenamefont
  {Luijten}(2023)}]{curk2023Phase}%
  \BibitemOpen
  \bibfield  {author} {\bibinfo {author} {\bibfnamefont {T.}~\bibnamefont
  {Curk}}\ and\ \bibinfo {author} {\bibfnamefont {E.}~\bibnamefont {Luijten}},\
  }\href {https://doi.org/10.48550/arXiv.2201.08922} {\bibinfo {title} {Phase
  {{Separation}} and {{Ripening}} in a {{Viscoelastic Gel}}}} (\bibinfo {year}
  {2023}),\ \Eprint {https://arxiv.org/abs/2201.08922} {arxiv:2201.08922
  [cond-mat]} \BibitemShut {NoStop}%
\bibitem [{\citenamefont {Zhang}\ \emph {et~al.}(2021)\citenamefont {Zhang},
  \citenamefont {Lee}, \citenamefont {Meir}, \citenamefont {Brangwynne},\ and\
  \citenamefont {Wingreen}}]{zhang2021Mechanical}%
  \BibitemOpen
  \bibfield  {author} {\bibinfo {author} {\bibfnamefont {Y.}~\bibnamefont
  {Zhang}}, \bibinfo {author} {\bibfnamefont {D.~S.~W.}\ \bibnamefont {Lee}},
  \bibinfo {author} {\bibfnamefont {Y.}~\bibnamefont {Meir}}, \bibinfo {author}
  {\bibfnamefont {C.~P.}\ \bibnamefont {Brangwynne}},\ and\ \bibinfo {author}
  {\bibfnamefont {N.~S.}\ \bibnamefont {Wingreen}},\ }\bibfield  {title}
  {\bibinfo {title} {Mechanical {{Frustration}} of {{Phase Separation}} in the
  {{Cell Nucleus}} by {{Chromatin}}},\ }\href
  {https://doi.org/10.1103/PhysRevLett.126.258102} {\bibfield  {journal}
  {\bibinfo  {journal} {Phys. Rev. Lett.}\ }\textbf {\bibinfo {volume} {126}},\
  \bibinfo {pages} {258102} (\bibinfo {year} {2021})}\BibitemShut {NoStop}%
\bibitem [{\citenamefont {Ohta}\ and\ \citenamefont
  {Kawasaki}(1986)}]{ohta1986Equilibrium}%
  \BibitemOpen
  \bibfield  {author} {\bibinfo {author} {\bibfnamefont {T.}~\bibnamefont
  {Ohta}}\ and\ \bibinfo {author} {\bibfnamefont {K.}~\bibnamefont
  {Kawasaki}},\ }\bibfield  {title} {\bibinfo {title} {Equilibrium morphology
  of block copolymer melts},\ }\href {https://doi.org/10.1021/ma00164a028}
  {\bibfield  {journal} {\bibinfo  {journal} {Macromolecules}\ }\textbf
  {\bibinfo {volume} {19}},\ \bibinfo {pages} {2621} (\bibinfo {year}
  {1986})}\BibitemShut {NoStop}%
\bibitem [{\citenamefont {Muratov}(2002)}]{muratov2002Theory}%
  \BibitemOpen
  \bibfield  {author} {\bibinfo {author} {\bibfnamefont {C.~B.}\ \bibnamefont
  {Muratov}},\ }\bibfield  {title} {\bibinfo {title} {Theory of domain patterns
  in systems with long-range interactions of {{Coulomb}} type},\ }\href
  {https://doi.org/10.1103/PhysRevE.66.066108} {\bibfield  {journal} {\bibinfo
  {journal} {Phys. Rev. E}\ }\textbf {\bibinfo {volume} {66}},\ \bibinfo
  {pages} {066108} (\bibinfo {year} {2002})}\BibitemShut {NoStop}%
\bibitem [{\citenamefont {Ren}\ and\ \citenamefont
  {Truskinovsky}(2000)}]{ren2000Finite}%
  \BibitemOpen
  \bibfield  {author} {\bibinfo {author} {\bibfnamefont {X.}~\bibnamefont
  {Ren}}\ and\ \bibinfo {author} {\bibfnamefont {L.}~\bibnamefont
  {Truskinovsky}},\ }\bibfield  {title} {\bibinfo {title} {Finite {{Scale
  Microstructures}} in {{Nonlocal Elasticity}}},\ }in\ \href
  {https://doi.org/10.1007/978-94-010-0728-3\_18} {\emph {\bibinfo {booktitle}
  {Advances in {{Continuum Mechanics}} and {{Thermodynamics}} of {{Material
  Behavior}}}}},\ \bibinfo {editor} {edited by\ \bibinfo {editor}
  {\bibfnamefont {D.~E.}\ \bibnamefont {Carlson}}\ and\ \bibinfo {editor}
  {\bibfnamefont {Y.-C.}\ \bibnamefont {Chen}}}\ (\bibinfo  {publisher}
  {{Springer Netherlands}},\ \bibinfo {address} {{Dordrecht}},\ \bibinfo {year}
  {2000})\ pp.\ \bibinfo {pages} {319--355}\BibitemShut {NoStop}%
\bibitem [{\citenamefont {Paulin}\ \emph {et~al.}(2022)\citenamefont {Paulin},
  \citenamefont {Morrow}, \citenamefont {Hennessy},\ and\ \citenamefont
  {MacMinn}}]{paulin2022Fluid}%
  \BibitemOpen
  \bibfield  {author} {\bibinfo {author} {\bibfnamefont {O.~W.}\ \bibnamefont
  {Paulin}}, \bibinfo {author} {\bibfnamefont {L.~C.}\ \bibnamefont {Morrow}},
  \bibinfo {author} {\bibfnamefont {M.~G.}\ \bibnamefont {Hennessy}},\ and\
  \bibinfo {author} {\bibfnamefont {C.~W.}\ \bibnamefont {MacMinn}},\
  }\bibfield  {title} {\bibinfo {title} {Fluid\textendash fluid phase
  separation in a soft porous medium},\ }\href
  {https://doi.org/10.1016/j.jmps.2022.104892} {\bibfield  {journal} {\bibinfo
  {journal} {J. Mech. Phys. Solids}\ }\textbf {\bibinfo {volume} {164}},\
  \bibinfo {pages} {104892} (\bibinfo {year} {2022})}\BibitemShut {NoStop}%
\bibitem [{\citenamefont {Heyden}\ \emph {et~al.}(2015)\citenamefont {Heyden},
  \citenamefont {Conti},\ and\ \citenamefont {Ortiz}}]{heyden2015Nonlocal}%
  \BibitemOpen
  \bibfield  {author} {\bibinfo {author} {\bibfnamefont {S.}~\bibnamefont
  {Heyden}}, \bibinfo {author} {\bibfnamefont {S.}~\bibnamefont {Conti}},\ and\
  \bibinfo {author} {\bibfnamefont {M.}~\bibnamefont {Ortiz}},\ }\bibfield
  {title} {\bibinfo {title} {A nonlocal model of fracture by crazing in
  polymers},\ }\href {https://doi.org/10.1016/j.mechmat.2015.02.006} {\bibfield
   {journal} {\bibinfo  {journal} {Mech. Mater.}\ }\textbf {\bibinfo {volume}
  {90}},\ \bibinfo {pages} {131} (\bibinfo {year} {2015})}\BibitemShut
  {NoStop}%
\bibitem [{\citenamefont {Nikolov}\ \emph {et~al.}(2007)\citenamefont
  {Nikolov}, \citenamefont {Han},\ and\ \citenamefont
  {Raabe}}]{nikolov2007Origin}%
  \BibitemOpen
  \bibfield  {author} {\bibinfo {author} {\bibfnamefont {S.}~\bibnamefont
  {Nikolov}}, \bibinfo {author} {\bibfnamefont {C.~S.}\ \bibnamefont {Han}},\
  and\ \bibinfo {author} {\bibfnamefont {D.}~\bibnamefont {Raabe}},\ }\bibfield
   {title} {\bibinfo {title} {On the origin of size effects in small-strain
  elasticity of solid polymers},\ }\href
  {https://doi.org/10.1016/j.ijsolstr.2006.06.039} {\bibfield  {journal}
  {\bibinfo  {journal} {Int. J. Solids Struct.}\ }\textbf {\bibinfo {volume}
  {44}},\ \bibinfo {pages} {1582} (\bibinfo {year} {2007})}\BibitemShut
  {NoStop}%
\bibitem [{\citenamefont {Bourdin}\ \emph {et~al.}(2008)\citenamefont
  {Bourdin}, \citenamefont {Francfort},\ and\ \citenamefont
  {Marigo}}]{bourdin2008Variational}%
  \BibitemOpen
  \bibfield  {author} {\bibinfo {author} {\bibfnamefont {B.}~\bibnamefont
  {Bourdin}}, \bibinfo {author} {\bibfnamefont {G.~A.}\ \bibnamefont
  {Francfort}},\ and\ \bibinfo {author} {\bibfnamefont {J.-J.}\ \bibnamefont
  {Marigo}},\ }\bibfield  {title} {\bibinfo {title} {The {{Variational
  Approach}} to {{Fracture}}},\ }\href
  {https://doi.org/10.1007/s10659-007-9107-3} {\bibfield  {journal} {\bibinfo
  {journal} {J. Elasticity}\ }\textbf {\bibinfo {volume} {91}},\ \bibinfo
  {pages} {5} (\bibinfo {year} {2008})}\BibitemShut {NoStop}%
\bibitem [{\citenamefont {Tanaka}(2022)}]{tanaka2022Viscoelastic}%
  \BibitemOpen
  \bibfield  {author} {\bibinfo {author} {\bibfnamefont {H.}~\bibnamefont
  {Tanaka}},\ }\bibfield  {title} {\bibinfo {title} {Viscoelastic phase
  separation in biological cells},\ }\href
  {https://doi.org/10.1038/s42005-022-00947-7} {\bibfield  {journal} {\bibinfo
  {journal} {Commun. Phys.}\ }\textbf {\bibinfo {volume} {5}},\ \bibinfo
  {pages} {1} (\bibinfo {year} {2022})}\BibitemShut {NoStop}%
\bibitem [{\citenamefont {Kim}\ \emph {et~al.}(2020)\citenamefont {Kim},
  \citenamefont {Liu}, \citenamefont {Weon}, \citenamefont {Cohen},
  \citenamefont {Hui}, \citenamefont {Dufresne},\ and\ \citenamefont
  {Style}}]{kim2020Extreme}%
  \BibitemOpen
  \bibfield  {author} {\bibinfo {author} {\bibfnamefont {J.~Y.}\ \bibnamefont
  {Kim}}, \bibinfo {author} {\bibfnamefont {Z.}~\bibnamefont {Liu}}, \bibinfo
  {author} {\bibfnamefont {B.~M.}\ \bibnamefont {Weon}}, \bibinfo {author}
  {\bibfnamefont {T.}~\bibnamefont {Cohen}}, \bibinfo {author} {\bibfnamefont
  {C.-Y.}\ \bibnamefont {Hui}}, \bibinfo {author} {\bibfnamefont {E.~R.}\
  \bibnamefont {Dufresne}},\ and\ \bibinfo {author} {\bibfnamefont {R.~W.}\
  \bibnamefont {Style}},\ }\bibfield  {title} {\bibinfo {title} {Extreme cavity
  expansion in soft solids: {{Damage}} without fracture},\ }\href
  {https://doi.org/10.1126/sciadv.aaz0418} {\bibfield  {journal} {\bibinfo
  {journal} {Sci. Adv.}\ }\textbf {\bibinfo {volume} {6}},\ \bibinfo {pages}
  {eaaz0418} (\bibinfo {year} {2020})}\BibitemShut {NoStop}%
\bibitem [{\citenamefont {{Raayai-Ardakani}}\ \emph {et~al.}(2019)\citenamefont
  {{Raayai-Ardakani}}, \citenamefont {Earl},\ and\ \citenamefont
  {Cohen}}]{raayai-ardakani2019Intimate}%
  \BibitemOpen
  \bibfield  {author} {\bibinfo {author} {\bibfnamefont {S.}~\bibnamefont
  {{Raayai-Ardakani}}}, \bibinfo {author} {\bibfnamefont {D.~R.}\ \bibnamefont
  {Earl}},\ and\ \bibinfo {author} {\bibfnamefont {T.}~\bibnamefont {Cohen}},\
  }\bibfield  {title} {\bibinfo {title} {The intimate relationship between
  cavitation and fracture},\ }\href {https://doi.org/10.1039/C9SM00570F}
  {\bibfield  {journal} {\bibinfo  {journal} {Soft Matter}\ }\textbf {\bibinfo
  {volume} {15}},\ \bibinfo {pages} {4999} (\bibinfo {year}
  {2019})}\BibitemShut {NoStop}%
\bibitem [{\citenamefont {{Vidal-Henriquez}}\ and\ \citenamefont
  {Zwicker}(2021)}]{vidal-henriquez2021Cavitation}%
  \BibitemOpen
  \bibfield  {author} {\bibinfo {author} {\bibfnamefont {E.}~\bibnamefont
  {{Vidal-Henriquez}}}\ and\ \bibinfo {author} {\bibfnamefont {D.}~\bibnamefont
  {Zwicker}},\ }\bibfield  {title} {\bibinfo {title} {Cavitation controls
  droplet sizes in elastic media},\ }\href
  {https://doi.org/10.1073/pnas.2102014118} {\bibfield  {journal} {\bibinfo
  {journal} {Proc. Natl. Acad. Sci. U.S.A.}\ }\textbf {\bibinfo {volume}
  {118}},\ \bibinfo {pages} {e2102014118} (\bibinfo {year} {2021})}\BibitemShut
  {NoStop}%
\bibitem [{\citenamefont {{Vidal-Henriquez}}\ and\ \citenamefont
  {Zwicker}(2020)}]{vidal-henriquez2020Theory}%
  \BibitemOpen
  \bibfield  {author} {\bibinfo {author} {\bibfnamefont {E.}~\bibnamefont
  {{Vidal-Henriquez}}}\ and\ \bibinfo {author} {\bibfnamefont {D.}~\bibnamefont
  {Zwicker}},\ }\bibfield  {title} {\bibinfo {title} {Theory of droplet
  ripening in stiffness gradients},\ }\href
  {https://doi.org/10.1039/D0SM00182A} {\bibfield  {journal} {\bibinfo
  {journal} {Soft Matter}\ }\textbf {\bibinfo {volume} {16}},\ \bibinfo {pages}
  {5898} (\bibinfo {year} {2020})}\BibitemShut {NoStop}%
\bibitem [{\citenamefont {Tyler}\ and\ \citenamefont
  {Morse}(2003)}]{tyler2003Stress}%
  \BibitemOpen
  \bibfield  {author} {\bibinfo {author} {\bibfnamefont {C.~A.}\ \bibnamefont
  {Tyler}}\ and\ \bibinfo {author} {\bibfnamefont {D.~C.}\ \bibnamefont
  {Morse}},\ }\bibfield  {title} {\bibinfo {title} {Stress in
  {{Self-Consistent-Field Theory}}},\ }\href
  {https://doi.org/10.1021/ma034601x} {\bibfield  {journal} {\bibinfo
  {journal} {Macromolecules}\ }\textbf {\bibinfo {volume} {36}},\ \bibinfo
  {pages} {8184} (\bibinfo {year} {2003})}\BibitemShut {NoStop}%
\bibitem [{\citenamefont {Thompson}\ \emph {et~al.}(2003)\citenamefont
  {Thompson}, \citenamefont {Rasmussen},\ and\ \citenamefont
  {Lookman}}]{thompson2003Improved}%
  \BibitemOpen
  \bibfield  {author} {\bibinfo {author} {\bibfnamefont {R.~B.}\ \bibnamefont
  {Thompson}}, \bibinfo {author} {\bibfnamefont {K.~O.}\ \bibnamefont
  {Rasmussen}},\ and\ \bibinfo {author} {\bibfnamefont {T.}~\bibnamefont
  {Lookman}},\ }\bibfield  {title} {\bibinfo {title} {Improved convergence in
  block copolymer self-consistent field theory by {{Anderson}} mixing},\ }\href
  {https://doi.org/10.1063/1.1629673} {\bibfield  {journal} {\bibinfo
  {journal} {J. Chem. Phys.}\ }\textbf {\bibinfo {volume} {120}},\ \bibinfo
  {pages} {31} (\bibinfo {year} {2003})}\BibitemShut {NoStop}%
\bibitem [{\citenamefont {Arora}\ \emph {et~al.}(2017)\citenamefont {Arora},
  \citenamefont {Morse}, \citenamefont {Bates},\ and\ \citenamefont
  {Dorfman}}]{arora2017Accelerating}%
  \BibitemOpen
  \bibfield  {author} {\bibinfo {author} {\bibfnamefont {A.}~\bibnamefont
  {Arora}}, \bibinfo {author} {\bibfnamefont {D.~C.}\ \bibnamefont {Morse}},
  \bibinfo {author} {\bibfnamefont {F.~S.}\ \bibnamefont {Bates}},\ and\
  \bibinfo {author} {\bibfnamefont {K.~D.}\ \bibnamefont {Dorfman}},\
  }\bibfield  {title} {\bibinfo {title} {Accelerating self-consistent field
  theory of block polymers in a variable unit cell},\ }\href
  {https://doi.org/10.1063/1.4986643} {\bibfield  {journal} {\bibinfo
  {journal} {J. Chem. Phys.}\ }\textbf {\bibinfo {volume} {146}},\ \bibinfo
  {pages} {244902} (\bibinfo {year} {2017})}\BibitemShut {NoStop}%
\end{thebibliography}%

\onecolumngrid
\appendix

\setcounter{figure}{0}
\renewcommand\thefigure{S\arabic{figure}}
\renewcommand{\theHfigure}{S\arabic{figure}}

\section{General local free energies cannot exhibit equilibrium patterns}
\label{si:local_cannot}

To show that local elasticity cannot yield patterns, we first consider a generic procedure to minimize the free energy functional, and explain afterwards that any periodic structure with finite period cannot be the minimum of the free energy functional if the interfacial term is the only nonlocal term.
Consider a free energy functional of a system of arbitrary dimension, which includes a volume-fraction-dependent term $\FF[\phi_i]$, an interfacial energy term $\FI[\phi_i]$ to penalize sharp interface, an elastic energy term $\FE[\boldu]$ and the constraint $\FC[\phi_i, \boldu, \zeta, \eta]$,
\begin{align}
     & F[\phi_i, \boldu, \zeta, \eta]
    = \FF[\phi_i]+\FI[\phi_i]+\FE[\boldu] + \FC[\phi_i, \boldu, \zeta, \eta] \;,
    \label{S-eqn:fe_functiuonal_generic}
\end{align}
where $\phi_i$ with $ i=1,2,\ldots,N$ are the volume fraction fields of $N$ components, $\boldu$ is the deformation vector field of the elastic component, and $\zeta$ and $\eta$ are two Lagrangian multipliers.
We keep the generic form of \Eqref{S-eqn:fe_functiuonal_generic} for simplicity and universality, except for the constraint, where we use
\begin{align}
     & \FC[\phi_i, \boldu, \zeta, \eta]
    = \int \bdx \zeta\biggl(\sum_i \phi_i - 1\biggr)+ \int \bdx \eta (J \phi_N - \phi_{N,0}) \;.
    \label{S-eqn:f_constraint}
\end{align}
The first term accounts for incompressibility, while the second one indicates that the $N$-th component is the elastic component, so its volume fraction is related to the displacement field by volume conservation.
Here, $\phi_{N,0}$ is the volume fraction distribution of the elastic component in the relaxed state (\figref{M-fig:schematic_nonlocal_d}A), and $J$ is the determinant of the deformation gradient tensor $\boldnabla_{\boldX} \boldu$, where $\boldX$ is the coordinate in the reference frame.
Extremizing the free energy in \Eqref{S-eqn:fe_functiuonal_generic} with respect to all of its variables leads to the corresponding self-consistent equations,
\begin{subequations}
    \label{S-eqn:scf_generic}
    \begin{align}
         & \frac{\delta \FF}{\delta \phi_i} + \frac{\delta \FI}{\delta \phi_i} + \zeta + \eta J \delta_{i,N} = 0 \\
         & \frac{\delta \FE}{\delta \boldu}+\frac{\delta }{\delta \boldu}\biggl(\int \bdx \eta J \phi_N\biggr)=0 \\
         & 1 - \sum_i \phi_i =0                                                                                  \\
         & J \phi_N - \phi_{N,0}=0 \;,
    \end{align}
\end{subequations}
where $\delta_{i,N}$ is the Kronecker delta.
Note that the last two equations are simply the incompressibility and the volume conservation of the elastic component, respectively.
The constraint term $\FC[\phi_i, \boldu, \zeta, \eta]$ does not contribute to the free energy when the constraints are satisfied.

To minimize the free energy, we follow two steps:
First we require that the phase separated structure is periodic, and describe the unit cell by a group of parameters $\boldtheta$ \cite{tyler2003Stress}.
For a given $\boldtheta$, \Eqsref{S-eqn:scf_generic} can be solved to obtain the \emph{free energy minimum with fixed unit cell} $F^*(\boldtheta)=F[\phi_i^*(\boldtheta), \boldu^*(\boldtheta), \zeta^*(\boldtheta), \eta^*(\boldtheta)]$, where the symbols with asterisk denote the solution of \Eqsref{S-eqn:scf_generic}.
The minimum of the free energy can then be obtained by optimizing $F^*$ with respect to the period $\boldtheta$.
Note that $F^*$ is not a functional, but a function of $\boldtheta$ instead.

Follow the steps in \cite{tyler2003Stress}, we find that the derivative of $F^*(\boldtheta)$ can be obtained from the partial derivative of the free energy functional with respect to the period $\boldtheta$ while keeping the shape of all the spatial functions unchanged,
\begin{align}
    \frac{\dd F^*}{\dd \boldtheta}
     & = \left.\frac{\partial \FF}{\partial \boldtheta}\right\vert_* + \left.\frac{\partial \FI}{\partial \boldtheta}\right\vert_* + \left. \frac{\partial \FE}{\partial \boldtheta}\right\vert_* \;.
    \label{S-eqn:fe_derivative_original}
\end{align}
Assuming the total volume of the system $V$ is constant, the total free energy can be replaced by the average free energy density $\bar{f} = F / V$,
\begin{align}
    \frac{\dd \bar{f}^*}{\dd \boldtheta}
     & = \left.\frac{\partial \barfF}{\partial \boldtheta}\right\vert_* +\left.\frac{\partial \barfI}{\partial \boldtheta}\right\vert_* + \left. \frac{\partial \barfE}{\partial \boldtheta}\right\vert_* \;.
    \label{S-eqn:fe_derivative}
\end{align}
Note that \Eqref{S-eqn:fe_derivative} does not impose any assumptions on the exact form of the terms.
For any local volume-fraction-dependent term $\FF[\phi_i]$, e.g., the Flory-Huggins free energy, the partial derivative with respect to $\boldtheta$ vanishes, since the variables $\phi_i$ are dimensionless and have no explicit dependence on the pattern length scale.
For any local elasticity, including nonlinear ones with large deformations, the elastic energy $\FE$ takes a local form of the deformation gradient tensor $\boldnabla_{\boldX} \boldu$, which is also dimensionless and has no explicit dependence on the length scale.
Therefore, the partial derivative of the local elastic energy vanishes as well, leading to
\begin{align}
    \frac{\dd \bar{f}_{\mathrm{local}}^*}{\dd \boldtheta}
     & = \left.\frac{\partial \barfI}{\partial \boldtheta}\right\vert_* \;.
    \label{S-eqn:fe_derivative_local}
\end{align}
Here, term $\left.\partial \barfI/\partial \boldtheta\right\vert_*$ does not vanish since the interfacial term usually depends on the gradient of the volume fraction fields $\boldnabla \phi_i$, which has the dimension of length$^{-1}$ and contains explicit dependence on the length scale.

\begin{figure}[t]
    \begin{center}
        \includegraphics[width = 0.5 \linewidth]{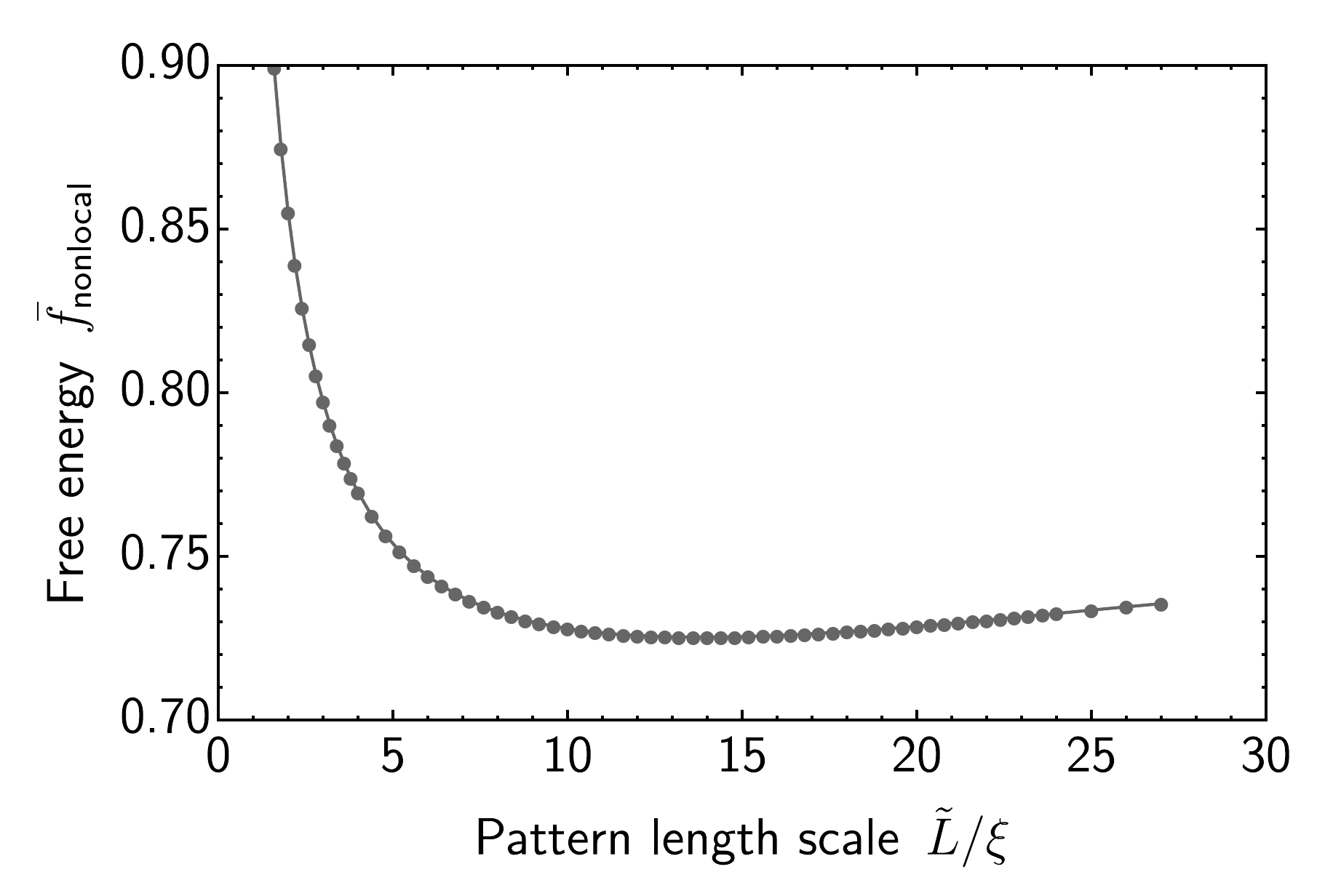}
    \end{center}
    \caption{
    \textbf{Free energy exhibits a minimum when varying pattern length scale.}
    Average free energy density $\bar{f}_{\mathrm{nonlocal}}=F_{\mathrm{nonlocal}}/V$ in units of $\kBT/\volS$ as a function of the length scale~$\tilde L$ of the periodic pattern obtained numerically.
    Model parameters are $E=0.02\kBT/\volS$, $\kappa =0.05\xi^2$, $\phi_0=1$, $\bar{\phiE}=0.5$, and $\chi=4$.
    }
    \label{S-fig:free_energy_curve}
\end{figure}

The period $\boldtheta$ typically contains not only the size but also the shape of the unit cell, including the angles between the base vectors of the unit cell.
Keeping the shape of the unit cell unchanged and denoting its size by $\tilde L$, we have
\begin{align}
     & \frac{\dd \bar{f}_{\mathrm{local}}^*}{\dd \tilde L} = \left.\frac{\partial \barfI}{\partial \tilde L}\right\vert_* \; ,
    \label{S-eqn:fe_derivative_local_size}
\end{align}
which is usually \emph{negative} since the interfacial energy prefers larger structure size.
Specifically, for the interfacial energy
\begin{align}
    \barfI = \frac{\kBT}{\volS} \frac{1}{V} \int \sum_i \frac{1}{2} \kappa_i (\boldnabla \phi_i)^2 \bdx \;,
\end{align}
where $\kappa_i$ quantifies the interfacial cost for component $i$, we have
\begin{align}
     & \left.\frac{\partial \barfI}{\partial \tilde L}\right\vert_* = -\frac{2}{\tilde L}\barfI^* = - \frac{\kBT}{\volS} \frac{2}{\tilde L} \frac{1}{V} \int \sum_i \frac{1}{2} \kappa_i (\boldnabla \phi_i^*)^2 \bdx \; ,
    \label{S-eqn:fe_derivative_interfacial}
\end{align}
which is negative.
Consequently, \Eqref{S-eqn:fe_derivative_local} states that with local elasticity, even if a periodic patterned structure is formed, it still cannot be the equilibrium state.
Instead, Ostwald ripening is inevitable, since the free energy can always be lowered by coarsening, and the equilibrium length scale diverges.
With nonlocal elasticity, the elastic energy will no longer be a local function of the deformation gradient tensor, implying two non-vanishing terms in \Eqref{S-eqn:fe_derivative}, which compete with each other,
\begin{align}
    \frac{\dd \bar{f}_{\mathrm{nonlocal}}^*}{\dd \tilde{L}}
     & = \left.\frac{\partial \barfI}{\partial \tilde{L}}\right\vert_* + \left. \frac{\partial \barfE}{\partial \tilde{L}}\right\vert_* \;,
    \label{S-eqn:fe_derivative_nonlocal}
\end{align}
leading to equilibrium pattern length scale (\figref{S-fig:free_energy_curve}).

The conclusion drawn from the free energy derivative above, although not a strict proof, can be understand in a more intuitive way:
Imagine a periodic patterned structure with average free energy $\bar{f}$, which is first scaled to be slightly larger, and then relaxed.
With local elasticity, the free energy decreases during scaling since only the interfacial term is affected.
Since the free energy must not be higher after relaxation due to the variational principle, the new free energy is generally lower than $\bar{f}$.
This process can be repeated  until the length diverges, implying local elasticity cannot explain the finite equilibrium length scale in a continuous field theory.

\section{Generic model of nonlocal elasticity}
\label{si:nonlocal}
The simplest (linear) nonlocal elasticity can be introduced by treating the elastic energy as a functional of the displacement field and expand it to second order \cite{kunin1982Elastic}.
We here focus on a one-dimensional system, and assume that the elastic energy $\FE$ is a functional of the displacement field $u(X)$, where $X$ is the coordinates in the reference frame.
Expanding $\FE$ to the second order, we have
\begin{align}
    \FE = \varPhi_0 + \int \dd X \varPhi_1(X)u(X) + \frac{1}{2}\int \dd X \dd X' \varPhi_2(X,X')u(X)u(X') + \cdots \;,
    \label{S-eqn:f_elastic_expand}
\end{align}
where $\varPhi_i$ are the derivatives of order $i$ of $\FE$.
Since the constant part is irrelevant and the expansion is made near the relaxed state, the $\varPhi_0$ and $\varPhi_1$ terms can be dropped.
Consequently, the lowest order approximation of $\FE$ reads
\begin{align}
    \FE \approx \frac{1}{2}\int \dd X \dd X' \varPhi(X,X')u(X)u(X') \;.
    \label{S-eqn:f_elastic_expand_lite}
\end{align}
Since the elastic energy must be invariant under constant shifts of the displacement field $u$,
\begin{align}
    \int \dd X' \varPhi(X,X')=0 \;.
    \label{S-eqn:f_elastic_Phi_ts_u}
\end{align}
Extracting the diagonal element from $\varPhi(X,X')$,
\begin{align}
    \varPhi(X,X') = \psi(X)\delta(X-X') - \varPsi(X,X') \;,
    \label{S-eqn:f_elastic_Psi_def}
\end{align}
and integrating both side with respect to $X'$ and using \Eqref{S-eqn:f_elastic_Phi_ts_u}, we find
\begin{align}
    \psi(X) = \int \dd X' \varPsi(X,X') \;.
    \label{S-eqn:f_elastic_Psi_property}
\end{align}
Note that the function $\varPsi$ is arbitrary and not subject to a constraint similar to \Eqref{S-eqn:f_elastic_Phi_ts_u}, in contrast to $\varPhi$.
Inserting \Eqref{S-eqn:f_elastic_Psi_def} into \Eqref{S-eqn:f_elastic_expand_lite} and using \Eqref{S-eqn:f_elastic_Phi_ts_u} again, we find
\begin{align}
    \FE \approx \frac{1}{4}\int \dd X \dd X' \varPsi(X,X')\bigl[u(X)-u(X')\bigr]^2 \;.
    \label{S-eqn:f_elastic_spring_model}
\end{align}
The equation above has a very clear physical picture:
The system contains multiple springs, with both ends of each spring tied to $X$ and $X'$ in the reference frame, and stretched by $u(X)-u(X')$ after the deformation of the elastic component.
The elastic energy of the system is just the total potential energy of all the springs, while the function $\varPsi$ is related to the stiffness of the springs.
Defining the strain $\epsilon=\dd u/\dd X$, this can be written as
\begin{align}
    \FE
     & \approx \frac{1}{4}\int \dd X \dd X' \varPsi(X,X')\biggl[\int_X^{X'} \dd X^* \epsilon(X^*)\biggr]^2                                  \notag \\
     & =\frac{1}{4}\int \dd X \dd \gamma \varPsi(X-\gamma/2,X+\gamma/2)\biggl[\int_{X-\gamma/2}^{X+\gamma/2} \dd X^* \epsilon(X^*)\biggr]^2 \notag \\
     & =\frac{1}{2}\int \dd X \dd X' \epsilon(X) \epsilon(X') c(X,X') \;,
    \label{S-eqn:f_elastic_nonlocal}
\end{align}
with
\begin{align}
    c(X,X') = \frac{1}{2}\int \dd \gamma \dd X \varPsi(X-\gamma/2,X+\gamma/2) \Pi_\gamma(X-X_1^*) \Pi_\gamma(X-X_2^*)
    \;,
\end{align}
where $\Pi_\gamma(X)$ is the box function centered at 0 with width $\gamma$.
Assuming  no explicit dependence on position, we have
\begin{align}
    \varPhi(X,X') & =\varPhi(X-X')
                  &
    \varPsi(X,X') & =\varPsi(X-X')
                  & \text{and}     &  &
    c(X,X')       & =c(X-X') \;.
    \label{S-eqn:f_elastic_ts_X}
\end{align}
Note that this does not require a homogeneous deformation of the elastic component, but only that the property of the continuum is homogeneous.
Defining the nonlocal stress $\sigma_{\mathrm{nonlocal}}(X) = \int \dd X' \epsilon(X) c(X-X')$, \Eqref{S-eqn:f_elastic_nonlocal} becomes
\begin{align}
    \FE
     & \approx \frac{1}{2}\int \epsilon(X) \sigma_{\mathrm{nonlocal}}(X) \dd X \;,
    \label{S-eqn:f_elastic_nonlocal_final}
\end{align}
which is the elastic energy term in \Eqref{M-eqn:f_elastic} of the main text.

The nonlocal elasticity given by \Eqref{S-eqn:f_elastic_nonlocal_final} can also be obtained from a more explicit model, such as a full network model of polymer network \cite{wu1993Improved}.
Assuming that the network is formed by multiple Gaussian chains, and each polymer chain has $N$ monomers, while its ends are fixed at $X-\gamma/2$ and $X+\gamma/2$ in the reference frame, respectively, the total elastic energy of the network will be proportional to
\begin{align}
    \FE
     & \propto \int \dd X \dd N \dd \gamma k_N [u(X+\gamma/2)-u(X-\gamma/2)]^2 w(N,\gamma) \;,
    \label{S-eqn:f_elastic_polymer}
\end{align}
where $k_N$ is the stiffness of the entropic spring of a polymer chain of $N$ monomers, $w(N,\gamma)$ is the joint distribution of $N$ and $\gamma$ to consider the polydispersity of the polymer chain and the end-to-end distance distribution.
Note that originally the energy of a polymer strand should be written in the form of $k_N \bigl[\gamma + u(X+\gamma/2)-u(X-\gamma/2)\bigr]^2$.
However, one can easily show that it is equivalent to \Eqref{S-eqn:f_elastic_polymer} except for a constant shift.
Change the order of the integration,
\begin{align}
    \FE \propto \int \dd X \dd \gamma \biggl[\int \dd N w(N,\gamma) k_N\biggr] \bigl[u(X+\gamma/2)-u(X-\gamma/2)\bigr]^2 \;,
    \label{S-eqn:f_elastic_polymer_re}
\end{align}
which is just another form of \Eqref{S-eqn:f_elastic_spring_model}.

\section{Numerical methods}
\label{si:numerics}
We minimize the nonlocal free energy using an iterative numerical scheme, which makes use of simple mixing and the Anderson mixing method to improve numerical stability and convergence \cite{thompson2003Improved}.
A variable-cell algorithm further improves the performance of the numerical method~\cite{tyler2003Stress,arora2017Accelerating}.
It makes use of the free energy derivative in \Eqref{S-eqn:fe_derivative_nonlocal}, and provides a way to obtain the optimal volume fraction profile and the equilibrium \patNameAbbr{} length scale $L$ at the same time, thus greatly reducing computational costs.

We first express the Flory-Huggins part of the free energy in \Eqref{M-eqn:f_FH} in a symmetric form, with not only the volume fraction fields of the elastic component and the solvent, $\phiEsym$ and $\phiS$, but also their conjugated fields $\wE$ and $\wS$, respectively.
In one dimension, the average Flory-Huggins free energy at given period $\tilde{L}$ reads
\begin{align}
    \barfF= & \frac{\kBT}{\volS} \frac{1}{\tilde{L}}\int_0^{\tilde{L}} \dd x \left[(\chi \phiEsym\phiS - \wE\phiEsym - \wS\phiS) - \barphiEsym \log \QE- \barphiS \log \QS
        + \barphiEsym\log\barphiEsym + \barphiS\log\barphiS \right]\;,
    \label{S-eqn:fe_FH_numerical}
\end{align}
where the single molecular partition functions $\QE$ and $\QS$ are defined as
\begin{align}
    \QE & = \frac{1}{\tilde{L}} \int_0^{\tilde{L}} \dd x e^{-\wE}
        &
    \QS & = \frac{1}{\tilde{L}} \int_0^{\tilde{L}} \dd x e^{-\wS} \;.
    \label{S-eqn:fe_FH_numerical_Q}
\end{align}
Minimizing the free energy in \Eqref{S-eqn:fe_functiuonal_generic} with respect to $\wE$ and $\wS$ gives
\begin{align}
    \phiEsym & = \frac{\barphiEsym}{\QE}e^{-\wE}
             &
    \phiS    & = \frac{\barphiS}{\QS}e^{-\wS} \;.
    \label{S-eqn:fe_FH_numerical_phi}
\end{align}
Note that compared to \Eqref{M-eqn:f_FH}, \Eqref{S-eqn:fe_FH_numerical} does not change the minimum of the free energy functional, since \Eqref{M-eqn:f_FH} can be fully recovered by inserting \Eqref{S-eqn:fe_FH_numerical_Q} and \Eqref{S-eqn:fe_FH_numerical_phi} into \Eqref{S-eqn:fe_FH_numerical}.
This method brings two advantages:
First, the explicit logarithm terms of $\phiEsym$ and $\phiS$ are removed, circumventing the numerical difficulty related to negative volume fractions.
Second, the average volume fraction is automatically kept constant, since $\QE$ and $\QS$ act as normalization factors.
The interfacial term is also reinterpreted in a symmetric form,
\begin{align}
    \barfI= \frac{\kBT}{\volS} \frac{1}{\tilde{L}}\int_0^{\tilde{L}} \dx \biggl[\frac{\kappa}{2} \biggl(\frac{\dd}{\dd x} \phiEsym\biggr)^2 + \frac{\kappa}{2} \biggl(\frac{\dd}{\dd x} \phiS\biggr)^2\biggr] \;,
    \label{S-eqn:fe_I_numerical}
\end{align}
while the elastic energy term
\begin{align}
    \barfE = & \frac{E}{2\tilde{L}}  \int_0^{\tilde{L}_0} \dX \int_{-\infty}^{+\infty} \dXp \frac{\dd u}{\dd X} \frac{\dd u}{\dd X'} g_\xi(X'-X) \;.
    \label{S-eqn:f_elastic_numerical}
\end{align}
is expressed with the displacement field $u$, where $\tilde{L}_0$ is the period in the reference frame.
The Lagrangian multipliers must also be included in the numerical calculation to enforce the incompressibility and the volume conservation,
\begin{align}
     & \barfC
    = \frac{\kBT}{\volS} \frac{1}{\tilde{L}}\int_0^{\tilde{L}} \dx \zeta(\phiEsym+\phiS-1)+ \frac{\kBT}{\volS} \frac{1}{\tilde{L}}\int_0^{\tilde{L}} \dx \eta \bigl(J \phiEsym - \phi_{0}\bigr) \;,
    \label{S-eqn:f_constraint_numerical}
\end{align}
where $J=\dd u /\dd X + 1$.

In addition to \Eqref{S-eqn:fe_FH_numerical_phi}, other self-consistent equations for numerical calculation can be obtained by minimizing the sum of the four energy terms in \Eqref{S-eqn:fe_FH_numerical} and \Eqsref{S-eqn:fe_I_numerical}--\ref{S-eqn:f_constraint_numerical}, which reads
\begin{subequations}
    \label{S-eqn:sc_equation_numerical}
    \begin{align}
        \wE      & = \chi \phiS -\kappa \frac{\dd^2}{\dd x^2} \phiEsym + \zeta + \eta J                                                                                     \\
        \wS      & = \chi \phiEsym -\kappa \frac{\dd^2}{\dd x^2} \phiS + \zeta                                                                                              \\
        0        & = \frac{\dd}{\dd X} \biggl(\frac{\volS E}{\kBT} \int_{-\infty}^{+\infty} \dXp \frac{\dd u}{\dd X'} g_\xi(X'-X)\biggr) + \phi_0 \frac{\dd}{\dd x}(\eta J) \\
        1        & =\phiEsym+\phiS                                                                                                                                          \\
        \phi_{0} & =J \phiEsym      \;.
    \end{align}
\end{subequations}
Note that we transform all the coordinates to the reference frame, where the convolution exhibits a simple form and can be calculated efficiently with fast Fourier transforms (FFT).
In practice, we use the periodic alternative $u^*(X)=u(X)-(\phi_0/\bar{\phi}-1)X$ as the free variable since the displacement field $u(X)$ is not periodic.
Using $u$, $\wE$ and $\wS$ as the main variables, the following iteration scheme solves \Eqsref{S-eqn:sc_equation_numerical},
\begin{subequations}
    \label{S-eqn:iteration_numerical}
    \begin{align}
        J^{(i)}                & = \frac{\dd u^{(i)}}{\dd X} + 1                                                                                                                                                \\
        (\eta J)^{(i)}         & = - \frac{1}{\phi_0}\int \dd X \biggl[J^{(i)} \frac{\dd}{\dd X} \biggl(\frac{\volS E}{\kBT} \int_{-\infty}^{+\infty} \dXp \frac{\dd u^{(i)}}{\dd X'} g_\xi(X'-X)\biggr)\biggr] \\
        \QE^{(i)}              & = \frac{1}{\tilde{L}} \int_0^{\tilde{L}_0} \dd X J^{(i)} e^{-\wE^{(i)}}                                                                                                        \\
        \QS^{(i)}              & = \frac{1}{\tilde{L}} \int_0^{\tilde{L}_0} \dd X J^{(i)} e^{-\wS^{(i)}}                                                                                                        \\
        \phiEsym^{(i)}         & = \frac{\barphiEsym}{\QE^{(i)}}e^{-\wE^{(i)}}                                                                                                                                  \\
        \phiS^{(i)}            & = \frac{\barphiS}{\QS^{(i)}}e^{-\wS^{(i)}}                                                                                                                                     \\
        \zeta^{(i)}            & = \frac{1}{2}\biggl[\wE^{(i)}+\wS^{(i)}-\kappa \frac{\dd^2}{\dd x^2} \phiEsym^{(i)} -\kappa \frac{\dd^2}{\dd x^2} \phiS^{(i)} -(\eta J)^{(i)}\biggr]                           \\
        u^{(i,\mathrm{new})}   & = \int \dd X \biggl(\frac{\phi_0}{\phiEsym^{(i)}}-1\biggr)                                                                                                                     \\
        \wE^{(i,\mathrm{new})} & = \chi \phiS^{(i)} -\kappa \frac{\dd^2}{\dd x^2} \phiEsym^{(i)} + \zeta^{(i)} + (\eta J)^{(i)}                                                                                 \\
        \wS^{(i,\mathrm{new})} & = \chi \phiEsym^{(i)} -\kappa \frac{\dd^2}{\dd x^2} \phiS^{(i)} + \zeta^{(i)} \;.
    \end{align}
\end{subequations}
To improve numerical stability, a simple mixing method is used in most cases, where the differences between the new fields and the old ones are partially accepted,
\begin{subequations}
    \label{S-eqn:simple_mixing}
    \begin{align}
        u^{(i+1)}   & = u^{(i)}+\lambda_u\bigl(u^{(i,\mathrm{new})} - u^{(i)}\bigr)         \\
        \wE^{(i+1)} & = \wE^{(i)}+\lambda_w\bigl(\wE^{(i,\mathrm{new})}-\wE^{(i)}\bigr)     \\
        \wS^{(i+1)} & = \wS^{(i)}+\lambda_w\bigl(\wS^{(i,\mathrm{new})}-\wS^{(i)}\bigr) \;.
    \end{align}
\end{subequations}
Here $\lambda_u$ and $\lambda_w$ are two empirical constants, which usually take value smaller than $0.1$.
To accelerate the convergence, Anderson mixing is also used every few steps \cite{thompson2003Improved}.

The variable-cell method is used to simultaneously optimize the period $\tilde{L}$ of the structure  during  iteration.
$\tilde{L}$ is evolved in the direction of lowering the free energy \cite{tyler2003Stress,arora2017Accelerating},
\begin{align}
    \tilde{L}^{(i+1)} = \tilde{L}^{(i)}-\lambda_{\tilde{L}} \frac{\volS \xi^2}{\kBT}  \biggl[\biggl(\frac{\partial \barfI}{\partial \tilde{L}}\biggr)^{(i)} + \biggl(\frac{\partial \barfE}{\partial \tilde{L}}\biggr)^{(i)}\biggr] \;,
    \label{S-eqn:variable_cell}
\end{align}
where the partial derivative of the interfacial energy and the elastic energy can be calculated by
\begin{subequations}
    \label{S-eqn:variable_cell_detail}
    \begin{align}
        \biggl(\frac{\partial \barfI}{\partial \tilde{L}}\biggr)^{(i)} & = -\frac{2}{\tilde{L}^{(i)}} \barfI^{(i)}                                                                                                                 \\
        \biggl(\frac{\partial \barfE}{\partial \tilde{L}}\biggr)^{(i)} & = \frac{E}{2\tilde{L}^{(i)}}  \int_0^{\tilde{L}_0} \dX \int_{-\infty}^{+\infty} \dXp \frac{\dd u^{(i)}}{\dd X} \frac{\dd u^{(i)}}{\dd X'} h_\xi(X'-X) \;,
    \end{align}
\end{subequations}
with the new kernel $h_\xi(X)=g_\xi(X)+X \dd g_\xi(X)/\dd X$.
The evolution of $\tilde{L}$ is also accelerated by Anderson mixing~\cite{arora2017Accelerating}.
When converged, $\tilde{L}$ reaches the equilibrium length scale $L$.

For all of the numerical results, we use periodic boundary condition and 2048 spatial sample points per period of the \patName{}.
The incompressibility and the relative square-mean-root of the field error are converged to below $10^{-5}$, while the free energy derivative with respect to the period is converged to below $10^{-8}$.

\begin{figure}
    \begin{center}
        \includegraphics[width = 1.0 \linewidth]{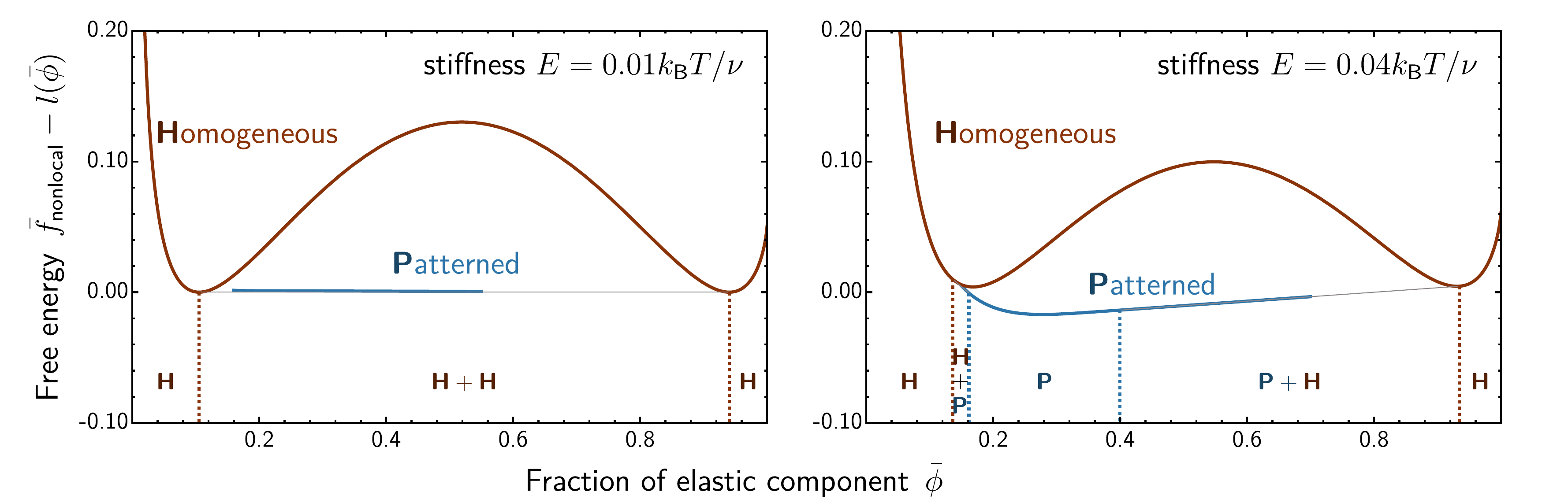}
    \end{center}
    \caption{
        \textbf{Common-tangent construction of free energies reveals coexistence states.}
        Average free energy density of \homoName{} (brown) and \patName{} (blue) as a function of the average fraction~$\bar\phi$ of the elastic component for a soft ($E = 0.01 \kBT / \volS$, left panel) and stiff system ($E = 0.04 \kBT/\volS$, right panel).
        The gray dashed lines mark common tangent lines.
        $l(\bar{\phi})$ is a linear function of $\bar{\phi}$ chosen for better visualization without affecting the common-tangent construction.
        Data obtained from full numerics at $\chi=3.2$, $\kappa \xi^{-2}=0.05$, and $\phi_0=1$.
    }
    \label{S-fig:phase_diagram_1D_real}
\end{figure}

To perform common-tangent construction at fixed interaction strength $\chi$ and stiffness $E$, the free energy curve is numerically sampled with the interval of average fraction $\bar{\phi}$ no larger than $0.01$.
Then, the numerical sample points are interpolated for the common-tangent construction (\figref{S-fig:phase_diagram_1D_real}).
We notice that the free energy difference between the periodic \patName{} and its coexistence with \homoName{} is tiny, which might be related to the irregularity of the droplet placement in real systems.

\section{Identifying the continuous phase transition}
\label{si:continuous}
We present two ways to identify the continuous phase transition, based on (i) overlap of spinodal and binodal lines and (ii) on a higher-order analysis.

\begin{figure}[t]
    \begin{center}
        \includegraphics[width = 1.0 \linewidth]{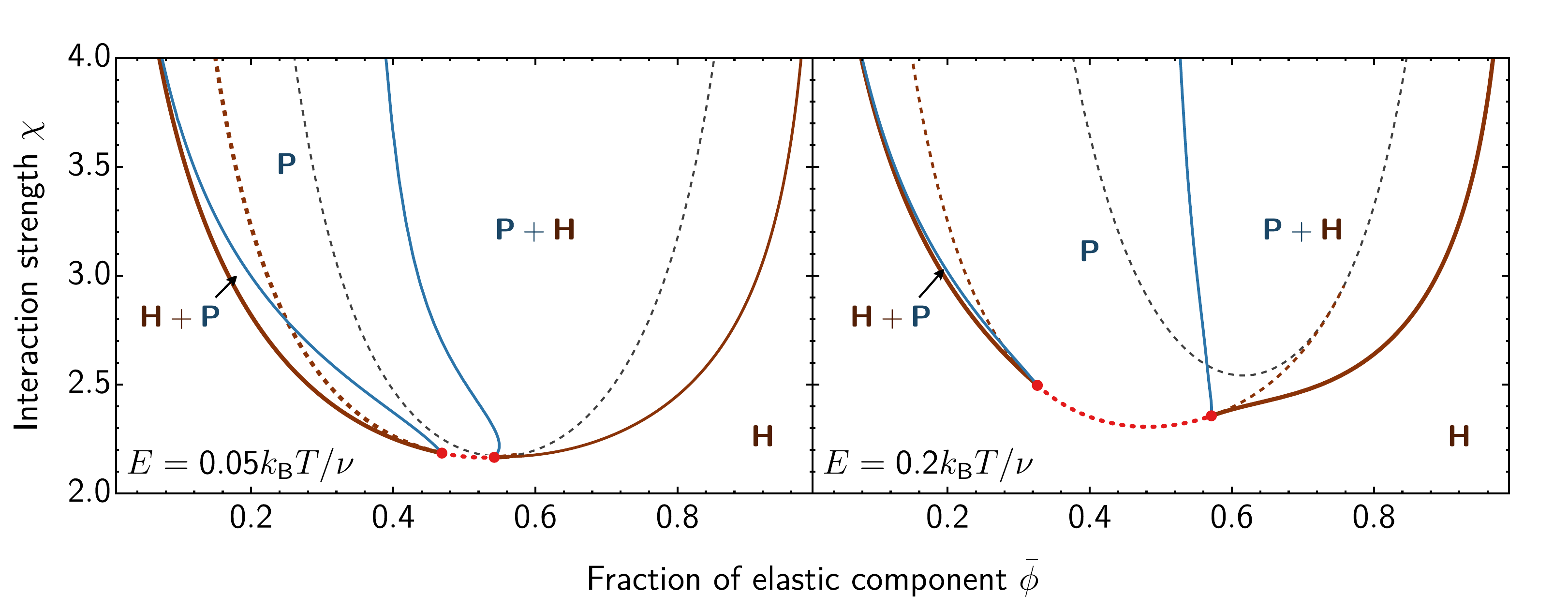}
    \end{center}
    \caption{
        \textbf{Continuous phase transition is identified in two different ways.}
        Phase diagram as a function of the average fraction $\bar\phi$ of the elastic component and interaction strength $\chi$ for two stiffnesses $E$.
        The brown and black dashed lines mark the spinodal curves considering the microphase separation and the macrophase separation, obtained by evaluating \Eqref{S-eqn:spinodal_micro} and \Eqref{S-eqn:spinodal_macro}, respectively.
        The binodal curves are obtained from the full numerics.
        The red dotted lines mark the window of continuous phase transition, obtained by considering the inequality in \Eqref{S-eqn:fe_multiple_modes_expansion_crit}.
    }
    \label{S-fig:phase_diagram_with_spinodal}
\end{figure}

\tocless\subsection{Spinodal and binodal overlap at continuous phase transition}
The spinodal line, based on linear stability analysis, and the binodal line, obtained from full numerics, overlap at the continuous phase transition.
In our system, we have multiple spinodals, since we can have  coexistence between two homogeneous and a patterned phase.

To get the spinodal of the homogeneous phase, we first substitute $\epsilon=J-1=\phi_0/\phiE-1$ into \Eqref{M-eqn:f_elastic} to express $\epsilon$ and $\sigma_{\mathrm{nonlocal}}$ with $J$, and transform the integral into the reference frame.
The average free energy density then reads
\begin{align}
    \bar{f} & = \frac{\kBT}{\volS} \frac{\bar{\phi}}{\phi_0}\frac{1}{\tilde{L}_0}\int_0^{\tilde{L}_0} \dX J \biggl[\phiE \log \phiE + \phiS\log(1-\phiE) + \chi\phiE(1-\phiE) + \kappa \biggl(\frac{\dd \phiE}{\dd X} \biggr)^2/J^2 \biggr] \notag \\
            & \quad + \frac{\bar{\phi}}{\phi_0}\frac{E}{2\tilde{L}_0} \int_0^{\tilde{L}_0} \dX \int_{-\infty}^{+\infty} \dXp \bigl[J(X)-1\bigr]\bigl[J(X')-1\bigr] g_\xi(X'-X) \;.
    \label{S-eqn:fe_full}
\end{align}
Since we have $J=\phi_0/\bar{\phi}$ for the homogeneous phase average volume fraction $\bar{\phi}$, we find
\begin{align}
    \bar{f}_{\mathrm{homo.}} & = \frac{\kBT}{\volS}\biggl(\bar{\phi} \log \bar{\phi} + (1-\bar{\phi})\log(1-\bar{\phi}) + \chi \bar{\phi}(1-\bar{\phi})\biggr) + \frac{E}{2} \frac{\bar{\phi}}{\phi_0}\biggl(\frac{\phi_0}{\bar{\phi}}-1\biggr)^2 \;.
    \label{S-eqn:fe_homo}
\end{align}
To test the stability, we first take the second-order derivative of the $\bar{f}_{\mathrm{homo.}}$ with respect to $\bar\phi$,
\begin{align}
    \frac{\partial^2 \bar{f}_{\mathrm{homo.}}}{\partial \bar{\phi}^2} = E \frac{\phi_0}{\bar{\phi}^3}+ \frac{\kBT}{\volS} \biggl(\frac{1}{1-\bar{\phi}}+\frac{1}{\bar{\phi}}-2 \chi \biggr) \;.
    \label{S-eqn:fe_homo_d2}
\end{align}
The spinodal of macrophase separation between two homogeneous phase corresponds to $\bar{f}_\mathrm{homo.}''(\bar\phi)=0$, yielding
\begin{align}
    \chi_{\mathrm{macro}} = \frac{1}{2}\biggl(\frac{\volS E}{\kBT} \frac{\phi_0}{\bar{\phi}^3}+\frac{1}{1-\bar{\phi}}+\frac{1}{\bar{\phi}}\biggr) \;;
    \label{S-eqn:spinodal_macro}
\end{align}
see the black dashed lines in \figref{S-fig:phase_diagram_with_spinodal}.

To consider microphase separation, we perturb $J$ from a constant value, $J=(\phi_0/\bar{\phi})\bigl(1+a \cos(q X)\bigr)$, where $a$ is the amplitude of the perturbation and $q$ is the associated wave number.
Evaluating \Eqref{S-eqn:fe_full}, taking the second-order derivative of $\bar{f}$ with respect to $a$, and taking the limit $a \to 0$, we find
\begin{align}
    \left.\frac{\partial^2 \bar{f}}{\partial a^2}\right\vert_{a=0} = E \frac{\phi_0 e^{-\frac{1}{8} \xi ^2 q^2}}{2 \bar{\phi}}+ \frac{\kBT}{\volS} \biggl(\frac{\kappa  q^2 \bar{\phi}^4}{\phi_0^2}-\chi  \bar{\phi}^2+\frac{\bar{\phi}^2}{2-2 \bar{\phi}}+\frac{\bar{\phi}}{2} \biggr) \;.
\end{align}
Stability of the homogeneous state requires $ \left.\partial^2 \bar{f}/\partial a^2\right\vert_{a=0}\ge0$ for all $q$, implying
\begin{align}
    \chi \le \chi_{\mathrm{u}} = \frac{\volS E}{\kBT} \frac{\phi_0 e^{-\frac{1}{8} \xi ^2 q^2}}{2 \bar{\phi}^3}+\frac{\kappa  q^2 \bar{\phi}^2}{\phi_0^2}+\frac{1}{2 \bar{\phi}}+\frac{1}{2-2 \bar{\phi}}
    \label{S-eqn:chi_requirement_for_given_q}
\end{align}
holds for all $q$.
$\chi_{\mathrm{u}}$ assumes its minimum at $q=q^*$ with
\begin{align}
    q^*=\frac{2 \sqrt{2}}{\xi } \sqrt{\log \biggl( \frac{\volS E}{\kBT} \frac{\xi ^2 \phi_0^3}{16 \kappa  \bar{\phi}^5} \biggr) } \;,
    \label{S-eqn:first_unstable_q}
\end{align}
if $(\volS E/\kBT) \xi ^2 \phi_0^3>16 \kappa  \bar{\phi}^5$, leading to the spinodal of microphase separation,
\begin{align}
    \chi_{\mathrm{micro}} = \frac{8 \kappa  \bar{\phi}^2 \log \bigl( \frac{\volS E}{\kBT} \frac{\xi ^2 \phi_0^3}{16 \kappa  \bar{\phi}^5} \bigr)}{\xi ^2 \phi_0^2}+\frac{8 \kappa  \bar{\phi}^2}{\xi ^2 \phi_0^2}+\frac{1}{2 \bar{\phi}}+\frac{1}{2-2 \bar{\phi}} \;;
    \label{S-eqn:spinodal_micro}
\end{align}
see the brown dashed lines in \figref{S-fig:phase_diagram_with_spinodal}.
Comparing this line to the binodal curve obtained from the full numerics, we find that they overlap in a parameter windows, which hints at the existence of a continuous phase transition (\figref{S-fig:phase_diagram_with_spinodal}).
Note that if $(\volS E/\kBT) \xi ^2 \phi_0^3>16 \kappa  \bar{\phi}^5$ does not hold then $\chi_{\mathrm{u}}$ takes its minimum at $q \to +0$, where $\chi_{\mathrm{u}}$ turns to $\chi_{\mathrm{macro}}$ and the macrophase spinodal given by \Eqref{S-eqn:spinodal_macro} is recovered.

\tocless\subsection{Higher-order analysis identifies the continuous phase transition}

\Eqref{S-eqn:spinodal_micro} defines the spinodal curve of the microphase separation in the $\phiE$-$\chi$ plane, which implies that the homogeneous phase is (meta-)stable below the curve, and unstable above the curve.
No information about stability is provided right \emph{on} the spinodal curve, since \Eqref{S-eqn:chi_requirement_for_given_q} is not a sufficient condition for stability when $\chi=\chi_\mathrm{u}$.In order to test the stability on the spinodal, we approximate $J$ as $J=(\phi_0/\bar{\phi})\bigl(1+ a_{q^*} \cos(q^* X) + \sum_{q \ne q^*} a_q \cos(q X)\bigr)$ and perturb the average volume fraction $\bar{\phi}$ as $\bar{\phi}+\delta\bar{\phi}$ at the same time.
The average free energy density can then be written as
\begin{align}
    \bar{f}=\bar{f}[\delta\bar{\phi}, a_{q^*}, a_{q_1}, a_{q_2},\ldots] \;.
    \label{S-eqn:fe_multiple_modes}
\end{align}
On the spinodal line of the homogeneous phase, we have
\begin{align}
    \left.\frac{\partial^2 \bar{f}}{(\partial \delta\bar{\phi})^2}\right\vert_{0} & >0      \quad\quad\quad\quad \left.\frac{\partial^2 \bar{f}}{(\partial a_{q^*})^2}\right\vert_{0}=0 \quad\quad\quad\quad \left.\frac{\partial^2 \bar{f}}{(\partial a_{q})^2}\right\vert_{0} >0 \quad \text{if} \quad q\ne q^* \;,
    \label{S-eqn:fe_multiple_modes_2nd}
\end{align}
while all second-order cross derivatives vanish.
Here $\left.\cdot\right\vert_{0}$ indicates the value at the homogeneous state on the spinodal.
Expanding the free energy up to fourth-order, we find
\begin{align}
    \bar{f}
     & = \left.\bar{f}\right\vert_{0} + \left.\frac{\partial \bar{f}}{\partial \delta\bar{\phi}}\right\vert_{0} \delta\bar{\phi}
    + \frac{1}{2}\left.\frac{\partial^2 \bar{f}}{(\partial \delta\bar{\phi})^2}\right\vert_{0} (\delta\bar{\phi})^2 + \frac{1}{2}\sum_{q \ne q^*}\left.\frac{\partial^2 \bar{f}}{(\partial a_{q})^2}\right\vert_{0} a_{q}^2
    \notag                                                                                                                                                                                                                                                             \\
     & \quad +\frac{1}{2}\left.\frac{\partial^3 \bar{f}}{(\partial a_{q^*})^2\partial \delta\bar{\phi}}\right\vert_{0} a_{q^*}^2\delta\bar{\phi} + \frac{1}{2}\left.\frac{\partial^3 \bar{f}}{(\partial a_{q^*})^2\partial a_{2q^*}}\right\vert_{0} a_{q^*}^2 a_{2q^*}
    +\frac{1}{24}\left.\frac{\partial^4 \bar{f}}{(\partial a_{q^*})^4}\right\vert_{0} a_{q^*}^4
    +o\bigl(\left\lVert (\delta\bar{\phi}, a_{q^*}^2, a_{q_1}, a_{q_2},\ldots) \right\rVert^2 \bigr) \;,
    \label{S-eqn:fe_multiple_modes_expansion}
\end{align}
where omitted terms are either zero or absorbed in the remainder.
The stability of the homogeneous phase on the spinodal then requires
\begin{align}
    \frac{1}{24}\left.\frac{\partial^4 \bar{f}}{(\partial a_{q^*})^4}\right\vert_{0}
    - {\biggl(\frac{1}{2}\left.\frac{\partial^3 \bar{f}}{(\partial a_{q^*})^2\partial \delta\bar{\phi}}\right\vert_{0}\biggr)^2}\bigg/\biggl({2\left.\frac{\partial^2 \bar{f}}{(\partial \delta\bar{\phi})^2}\right\vert_{0}}\biggr)
    - {\biggl(\frac{1}{2}\left.\frac{\partial^3 \bar{f}}{(\partial a_{q^*})^2\partial a_{2q^*}}\right\vert_{0}\biggr)^2}\bigg/\biggl({2\left.\frac{\partial^2 \bar{f}}{(\partial a_{2q^*})^2}\right\vert_{0}}\biggr)
    \ge 0 \;.
    \label{S-eqn:fe_multiple_modes_expansion_crit}
\end{align}
Solving this inequality for $\bar{\phi}$ numerically, we obtain the window of $\bar{\phi}$ with a continuous phase transition.
Combined with \Eqref{S-eqn:spinodal_micro}, the phase boundary of the continuous phase transition can be obtained, which is marked as red dotted lines or red surface in \figref{S-fig:phase_diagram_with_spinodal}, \figref{M-fig:phase_diagram_grandcanonical} and \figref{M-fig:phase_diagram}.
In fact, in all these phase diagrams, the continuous transitions are verified with both the overlapping of spinodal and binodal, as well as the higher-order stability analysis.

\vspace{8mm}
\section{Approximate model and asymptotic solutions}
\label{si:approx_model}

To understand the scaling law of the length scale~$L$, we assume sharp interfaces and approximate the volume fraction profile~$\phi(x)$ as  a box function,
\begin{align}
    \phiE(x)=\phi_{+} + (\phi_{-}-\phi_{+})\Pi_{\alpha L}(x-\tilde{L}/2) \;,
    \label{S-eqn:profile_box}
\end{align}
within one period $x\in[0,\tilde{L})$.
Here $\alpha=(\phi_{+}-\bar{\phi})/(\phi_{+}-\phi_{-})$ is the fraction of the solvent-rich region relative to the period $\tilde{L}$, $\bar{\phi}$ is the average volume fraction of the elastic component in the deformed state, and $\phi_{-}$ and $\phi_{+}$ are the minimum and the maximum value of the volume fraction profile, respectively.
Converting the profile to the reference frame and making use of the relationship between strain and the volume fraction given by \Eqref{M-eqn:strain_fraction_relation}, we find
\begin{align}
    \epsilon(X)=\frac{\phi_0}{\phi_{+}} - 1 + \biggl(\frac{\phi_0}{\phi_{-}}-\frac{\phi_0}{\phi_{+}}\biggr)\Pi_{\alpha_0 \tilde{L}_0}\biggl(X-\frac{\tilde{L}_0}{2}\biggr) \;,
    \label{S-eqn:profile_box_strain_rf}
\end{align}
where $\tilde{L}_0=(\bar{\phi}/\phi_0)\tilde{L}$ and $\alpha_0=(\phi_{-}/\bar{\phi})\alpha$ are the period and relative droplet size in the reference frame, respectively.

To evaluate the elastic energy, we first consider the case where the period is much larger than the microscopic length scale ($\tilde{L} \gg \xi$).
In this case, we can safely ignore the interference between the neighboring periods since the Gaussian convolution kernel decays exponentially with distance.
The elastic energy density thus reads
\begin{align}
    \label{S-eqn:fe_regimeBC_general}
    \barfE & = \frac{E}{2}\biggl(\frac{\phi_0}{\phi_{-}}-\frac{\phi_0}{\phi_{+}}\biggr)^2\biggl(\frac{\xi}{\tilde{L}}\frac{e^{-2\beta^2\tilde{L}^2/\xi^2}-1}{\sqrt{2\pi}}+\tilde{L} \, \erf \biggl(\sqrt{2}\beta \frac{\tilde{L}}{\xi}\biggr)\biggr) + \bar{f}_{\mathrm{el,0}}
           & \text{with}                                                                                                                                                                                                                                                       &  &
    \beta  & = \frac{\phi_{-}(\phi_{+}-\bar{\phi})}{\phi_0(\phi_{+}-\phi_{-})} \;.
\end{align}
where $\bar{f}_{\mathrm{el,0}}$ is a term with no dependence on $\tilde{L}$,
\begin{align}
    \bar{f}_{\mathrm{el,0}} & =\frac{E}{2}\biggl[2\biggl(1-\frac{\bar{\phi}}{\phi_0}\biggr)\epsilon_0-\frac{\bar{\phi}}{\phi_0}\epsilon_0^2\biggr] \;,
\end{align}
with $\epsilon_0=\phi_0/\phi_{+}-1$.
Expanding \Eqref{S-eqn:fe_regimeBC_general} around $\tilde{L}\to+0$, we have
\begin{align}
     & \barfE^{\,\RN{2}} = E\frac{(\phi_{+} - \bar{\phi})^2}{\sqrt{2\pi}\phi_{+}^2}\frac{\tilde{L}}{\xi} + \bar{f}_{\mathrm{el,0}} +o(\tilde{L}^3) \;,
    \label{S-eqn:fe_regimeB}
\end{align}
while expanding around $\tilde{L}\to+\infty$ leads to
\begin{align}
     & \barfE^{\,\RN{3}} = E\frac{(\phi_{+}-\bar{\phi})(\phi_{+}-\phi_{-})\phi_0}{\phi_{-}\phi_{+}^2} - E\frac{1}{2\sqrt{2\pi}}\biggl(\frac{\phi_0}{\phi_{-}}-\frac{\phi_0}{\phi_{+}}\biggr)^2\frac{\xi}{\tilde{L}} + \bar{f}_{\mathrm{el,0}} +o\biggl(\frac{1}{\tilde{L}^2}\biggr) \;.
    \label{S-eqn:fe_regimeC}
\end{align}
Next, we consider the case $\tilde{L}\ll\xi$, where the elastic energy can be derived from the physical picture.
Since the period is much smaller than the microscopic length scale $\xi$, the convolution of the strain field simply gives the average strain which is not affected by the period $\tilde{L}$.
Evaluating \Eqref{M-eqn:f_elastic} and converting it to energy density, we obtain
\begin{align}
     & \barfE^{\,\RN{1}} = \frac{E}{2}\frac{\phi_0}{\bar{\phi}}\biggl(\frac{\bar{\phi}}{\phi_0}-1\biggr)^2 \;.
    \label{S-eqn:fe_regimeA}
\end{align}
Combining the results in \Eqref{S-eqn:fe_regimeB}, \Eqref{S-eqn:fe_regimeC} and \Eqref{S-eqn:fe_regimeA}, we find
\begin{align}
     & \barfE \approx \begin{cases}
                          \dfrac{E}{2}\dfrac{\phi_0}{\bar{\phi}}\bigl(\dfrac{\bar{\phi}}{\phi_0}-1\bigr)^2                                                                                                                                    & \tilde{L}<L_\mathrm{min}                \\
                          \bar{f}_{\mathrm{el,0}} + E\dfrac{(\phi_{+}-\bar{\phi})^2}{\sqrt{2\pi}\phi_{+}^2}\dfrac{\tilde{L}}{\xi}                                                                                                             & L_\mathrm{min}<\tilde{L}<L_\mathrm{max} \\
                          \bar{f}_{\mathrm{el,0}} +E\dfrac{(\phi_{+}-\phi_{-})(\phi_{+}-\bar{\phi})\phi_0}{\phi_{-}\phi_{+}^2} - E\dfrac{1}{2\sqrt{2\pi}}\bigl(\frac{\phi_0}{\phi_{-}}-\dfrac{\phi_0}{\phi_{+}}\bigr)^2\dfrac{\xi}{\tilde{L}} & \tilde{L}>L_\mathrm{max}
                      \end{cases} \;,
    \label{S-eqn:fe_piecewise}
\end{align}
where the boundary values $L_\mathrm{min}$ and $L_\mathrm{max}$ will be estimated later.
Differentiating $\barfE$ with respect to $L$, we have
\begin{align}
     & \frac{\partial \barfE}{\partial \tilde{L}} \approx \begin{cases}
                                                              0                                                                                                               & \tilde{L}<L_\mathrm{min}                \\
                                                              E\dfrac{(\phi_{+}-\bar{\phi})^2}{\sqrt{2\pi}\phi_{+}^2}\dfrac{1}{\xi}                                           & L_\mathrm{min}<\tilde{L}<L_\mathrm{max} \\
                                                              E\dfrac{1}{2\sqrt{2\pi}}\bigl(\dfrac{\phi_0}{\phi_{-}}-\dfrac{\phi_0}{\phi_{+}}\bigr)^2\dfrac{\xi}{\tilde{L}^2} & \tilde{L}>L_\mathrm{max}
                                                          \end{cases}\; .
    \label{S-eqn:fe_derivatives_piecewise}
\end{align}
Since the derivatives of the average free energy density govern the equilibrium length scale (see \Eqref{S-eqn:fe_derivative_nonlocal}), we determine $L_\mathrm{max}$ by balancing the last two terms of the derivatives of $\barfE$, resulting in
\begin{align}
     & L_\mathrm{max}=\frac{1}{\sqrt{2}}\frac{\phi_0}{\phi_{-}}\frac{\phi_{+}-\phi_{-}}{\phi_{+}-\bar{\phi}}\xi
    \;.
    \label{S-eqn:length_bound_max}
\end{align}
In contrast, the derivatives are all constants in the first two regimes of \Eqref{S-eqn:fe_derivatives_piecewise}, so we cannot estimate the boundary in the same way.
We thus balance $\barfE$ directly in the first two regimes to get
\begin{align}
     & L_\mathrm{min}=\sqrt{\frac{\pi}{2}}\frac{\phi_0}{\bar{\phi}}\xi
    \;.
    \label{S-eqn:length_bound_min}
\end{align}
Converting the two bounds $L_\mathrm{min}$ and $L_\mathrm{max}$ to the reference frame yields \Eqref{M-eqn:length_bound_ref} in the main text.

For completeness, we here also present the generic expression of $\barfE$ of the approximated model with the $\vartheta$ function
\begin{align}
    \barfE
     & =\frac{E}{2\tilde{L}}\int_{(1-\alpha_0)\tilde{L}_0/2}^{(1+\alpha_0)\tilde{L}_0/2} \dX \dXp \biggl(\frac{\phi_0}{\phi_{-}}-\frac{\phi_0}{\phi_{+}}\biggr)^2 \vartheta_3\biggl(\frac{\pi(X'-X)}{\tilde{L}_0},e^{-\frac{\pi ^2 \xi ^2}{2 \tilde{L}_0^2}}\biggr) + \bar{f}_{\mathrm{el,0}} \;.
    \label{S-eqn:fe_theta_function}
\end{align}
This integral can be evaluated numerically; see the black line in \figref{M-fig:box_function_approx}C for its derivative.
Note that all orders of derivative with respect to $\tilde{L}$ at $\tilde{L}\to+0$ are zero, so the free energy is almost independent of $\tilde{L}$, consistent with our picture to derive \Eqref{S-eqn:fe_regimeA}.

\end{document}